\documentclass[useAMS,usenatbib]{mn2e}
\usepackage{times}
\usepackage{graphicx}
\usepackage{subfigure}
\usepackage{psfrag}
\usepackage{amsmath}

\title[Multimodal nested sampling]{Multimodal nested sampling:
an efficient and robust alternative to MCMC methods for astronomical data analysis}  
\author[Farhan Feroz and M.P.~Hobson]
  {Farhan Feroz\thanks{E-mail: f.feroz@mrao.cam.ac.uk} and   
   M.P.~Hobson\\
  Astrophysics Group,
      Cavendish Laboratory, JJ Thomson Avenue,
      Cambridge CB3 0HE, UK\\
}

\date{Accepted ---. Received ---; in original form \today}
\pubyear{2007}

\begin{document}
\maketitle

\begin{abstract}
In performing a Bayesian analysis of astronomical data, two difficult
problems often emerge. First, in estimating the parameters of some
model for the data, the resulting posterior distribution may be
multimodal or exhibit pronounced (curving) degeneracies, which can
cause problems for traditional Markov Chain Monte Carlo (MCMC)
sampling methods.  Second, in selecting between a set of competing
models, calculation of the Bayesian evidence for each model is
computationally expensive using existing methods such as thermodynamic
integration. The nested sampling method introduced by \citet{Skilling},
has greatly reduced the computational expense of calculating evidences
and also produces posterior inferences as a by-product. This method
has been applied successfully in cosmological applications by
\citet{Mukherjee}, but their implementation was efficient only for
unimodal distributions without pronounced degeneracies. \citet{Shaw}
recently introduced a clustered nested sampling method which is
significantly more efficient in sampling from multimodal posteriors
and also determines the expectation and variance of the final evidence
from a {\em single} run of the algorithm, hence providing a further
increase in efficiency. In this paper, we build on the work of Shaw et
al. and present three new methods for sampling and evidence evaluation
from distributions that may contain multiple modes and significant
degeneracies in very high dimensions; we also present an even more
efficient technique for estimating the uncertainty on the evaluated
evidence. These methods lead to a further substantial improvement in
sampling efficiency and robustness, and are applied to two toy
problems to demonstrate the accuracy and economy of the evidence
calculation and parameter estimation. Finally, we discuss the use of
these methods in performing Bayesian object detection in astronomical
datasets, and show that they significantly outperform existing MCMC
techniques. An implementation of our methods will be publicly released
shortly.
\end{abstract}


\section{Introduction}
\label{intro} 

Bayesian analysis methods are now widely used in astrophysics and
cosmology, and it is thus important to develop methods for performing
such analyses in an efficient and robust manner.  In general, Bayesian
inference divides into two categories: parameter estimation and model
selection. Bayesian parameter estimation has been used quite
extensively in a variety of astronomical applications, although
standard MCMC methods, such as the basic
Metropolis--Hastings algorithm or the Hamiltonian sampling
technique (see e.g. \citet{MacKay}), can experience problems in sampling
efficiently from a multimodal posterior distribution or one with large
(curving) degeneracies between parameters. Moreover, MCMC methods often require
careful tuning of the proposal distribution to sample efficiently, and testing
for convergence can be problematic. Bayesian model selection has been
hindered by the computational expense involved in the calculation to
sufficient precision of the key ingredient, the Bayesian evidence (also
called the marginalised likelihood or the marginal density of the
data).  As the average likelihood of a model over its prior
probability space, the evidence can be used to assign relative
probabilities to different models (for a review of cosmological
applications, see \citet{Mukherjee}).  The existing preferred
evidence evaluation method, again based on MCMC techniques, is
thermodynamic integration (see e.g. \citet{ORuanaidh}),
which is extremely computationally intensive but has been used
successfully in astronomical applications (see e.g. \citet{McLachlan}; 
\citet{Marshall}; \citet{Slosar}; \citet{Niarchou};
\citet{Basset}; \citet{Trotta}; \citet{Beltran}; \citet{Bridges}). 
Some fast approximate methods have been used for evidence
evaluation, such as treating the posterior as a multivariate
Gaussian centred at its peak (see e.g. \citet{Hobson}),
but this approximation is clearly a poor one for multimodal posteriors
(except perhaps if one performs a separate Gaussian approximation at
each mode). The Savage--Dickey density ratio has also been proposed
(Trotta 2005) as an exact, and potentially faster, means of evaluating
evidences, but is restricted to the special case of nested hypotheses
and a separable prior on the model parameters. Various alternative
information criteria for astrophysical model selection are discussed
by \citet{Liddle}, but the evidence remains the preferred method.

The nested sampling approach (Skilling 2004) is a Monte Carlo
method targetted at the efficient calculation of the
evidence, but also produces posterior inferences as a by-product. In
cosmological applications, \citet{Mukherjee} show that their
implementation of the method requires a factor of $\sim 100$ fewer
posterior evaluations than thermodynamic integration. To achieve an
improved acceptance ratio and efficiency, their algorithm uses an
elliptical bound containing the current point set at each stage of the
process to restrict the region around the posterior peak from which
new samples are drawn. \citet{Shaw} point out, however, that this
method becomes highly inefficient for multimodal posteriors, and hence
introduce the notion of clustered nested sampling, in which multiple
peaks in the posterior are detected and isolated, and separate
ellipsoidal bounds are constructed around each mode. This approach
significantly increases the sampling efficiency. The overall
computational load is reduced still further by the use of an improved
error calculation (Skilling 2004) on the final evidence result that
produces a mean and standard error in one sampling, eliminating the
need for multiple runs.

In this paper, we build on the work of \citet{Shaw}, by pursuing further
the notion of detecting and characterising multiple modes in the
posterior from the distribution of nested samples. In particular,
within the nested sampling paradigm, we suggest three new algorithms
(the first two based on sampling from ellipsoidal bounds and the third
on the Metropolis algorithm) for calculating the evidence from a
multimodal posterior with high accuracy and efficiency even when the
number of modes is unknown, and for producing reliable posterior
inferences in this case.  The first algorithm samples from all the
modes simultaneously and provides an efficient way of calculating the
`global' evidence, while the second and third algorithms retain the
notion from Shaw et al. of identifying each of the posterior modes and
then sampling from each separately.  As a result, these algorithms can
also calculate the `local' evidence associated with each mode as well
as the global evidence. All the algorithms presented differ from that
of Shaw et al. in several key ways. Most notably, the identification
of posterior modes is performed using the X-means clustering algorithm
(Pelleg et al. 2000), rather than $k$-means clustering with $k=2$; we
find this leads to a substantial improvement in sampling efficiency
and robustness for highly multimodal posteriors. Further innovations
include a new method for fast identification of overlapping
ellipsoidal bounds, and a scheme for sampling consistently from any
such overlap region. A simple modification of our methods also enables
efficient sampling from posteriors that possess pronounced
degeneracies between parameters. Finally, we also present a yet more
efficient method for estimating the uncertainty in the calculated
(local) evidence value(s) from a single run of the algorithm. The
above innovations mean our new methods constitute a viable, general
replacement for traditional MCMC sampling techniques in astronomical
data analysis.

The outline of the paper is as follows. In section 2, we briefly
review the basic aspects of Bayesian inference for parameter
estimation and model selection. In section 3 we introduce nested
sampling and discuss the ellipsoidal nested sampling technique in
section 4.  We present two new algorithms based on ellipsoidal
sampling and compare them with previous methods in section 5, and in
Section 6 we present a new method based on the Metropolis algorithm.
In section 7, we apply our new algorithms to two toy problems to
demonstrate the accuracy and efficiency of the evidence calculation
and parameter estimation as compared with other techniques. In section
8, we consider the use of our new algorithms in Bayesian object
detection. Finally, our conclusions are presented in Section 9.

\section{Bayesian Inference}
\label{bayesian_infer}

Bayesian inference methods provide a consistent approach to the
estimation of a set parameters $\mathbf{\Theta}$ in a model (or
hypothesis) $H$ for the data $\mathbf{D}$. Bayes' theorem states that
\begin{equation} \Pr(\mathbf{\Theta}|\mathbf{D}, H) =
\frac{\Pr(\mathbf{D}|\,\mathbf{\Theta},H)\Pr(\mathbf{\Theta}|H)}
{\Pr(\mathbf{D}|H)},
\end{equation}
where $\Pr(\mathbf{\Theta}|\mathbf{D}, H) \equiv P(\mathbf{\Theta})$ 
is the posterior probability
distribution of the parameters, $\Pr(\mathbf{D}|\mathbf{\Theta}, H)
\equiv L(\mathbf{\Theta})$ is the likelihood,
$\Pr(\mathbf{\Theta}|H) \equiv \pi(\mathbf{\Theta})$ is the prior, and
$\Pr(\mathbf{D}|H) \equiv \mathcal{Z}$ is the Bayesian evidence.  

In parameter estimation, the normalising evidence factor is usually
ignored, since it is independent of the parameters $\mathbf{\Theta}$,
and inferences are obtained by taking samples from the (unnormalised)
posterior using standard MCMC sampling methods, where at equilibrium
the chain contains a set of samples from the parameter space
distributed according to the posterior. This posterior constitutes the
complete Bayesian inference of the parameter values, and can be
marginalised over each parameter to obtain individual parameter
constraints.

In contrast to parameter estimation problems, in model selection the
evidence takes the central role and is simply the factor required to
normalize the posterior over $\mathbf{\Theta}$:
\begin{equation}
\mathcal{Z} =
\int{L(\mathbf{\Theta})\pi(\mathbf{\Theta})}d^D\mathbf{\Theta},
\label{eq:3}
\end{equation} 
where $D$ is the dimensionality of the parameter space.  As the
average of the likelihood over the prior, the evidence is larger for a
model if more of its parameter space is likely and smaller for a model
with large areas in its parameter space having low likelihood values,
even if the likelihood function is very highly peaked. Thus, the
evidence automatically implements Occam's razor: a simpler
theory with compact parameter space will have a larger evidence than a
more complicated one, unless the latter is significantly better at
explaining the data. 
The question of model selection between two
models $H_{0}$ and $H_{1}$ can then be decided by comparing their
respective posterior probabilities given the observed
data set $\mathbf{D}$, as follows
\begin{equation}
\frac{\Pr(H_{1}|\mathbf{D})}{\Pr(H_{0}|\mathbf{D})}
=\frac{\Pr(\mathbf{D}|H_{1})\Pr(H_{1})}{\Pr(\mathbf{D}|
H_{0})\Pr(H_{0})}
=\frac{\mathcal{Z}_1}{\mathcal{Z}_0}\frac{\Pr(H_{1})}{\Pr(H_{0})},
\label{eq:3.1}
\end{equation}
where $\Pr(H_{1})/\Pr(H_{0})$ is the a priori probability ratio for
the two models, which can often be set to unity but occasionally
requires further consideration.

Unfortunately, evaluation of the multidimensional integral
\eqref{eq:3} is a challenging numerical task.  The standard technique
is thermodynamic integration, which uses a modified form of MCMC
sampling.  The dependence of the evidence on the prior requires 
that the prior space is adequately sampled, even in regions
of low likelihood. To achieve this, the thermodynamic
integration technique draws MCMC samples not from the posterior
directly but from $L^{\lambda}\pi$ where $\lambda$ is an
inverse temperature that is raised from $\approx 0$ to $1$. For low
values of $\lambda$, peaks in the posterior are sufficiently suppressed
to allow improved mobility of the chain over the entire prior
range. Typically it is possible to obtain accuracies of within 0.5
units in log-evidence via this method, but in cosmological
applications it typically requires of order $10^6$ samples per chain
(with around 10 chains required to determine a sampling error). This
makes evidence evaluation at least an order of magnitude more costly
than parameter estimation.

\begin{figure}
\begin{center}
\includegraphics[width=1.0\columnwidth]{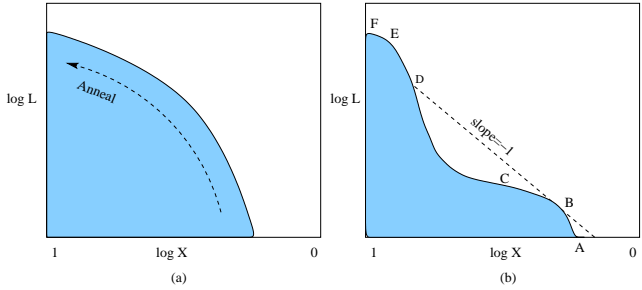}
\caption{Proper thermodynamic integration requires the log-likelihood to be
concave like (a), not (b).}
\label{fig:phase_chg}
\end{center}
\end{figure}

Another problem faced by thermodynamic integration is in navigating through
phase changes as pointed out by \citet{Skilling}. As $\lambda$ increases from 0
to 1, one hopes that the thermodynamic integration tracks gradually up in $L$ so
inwards in $X$ as illustrated in Fig.~\ref{fig:phase_chg}(a). $\lambda$ is
related to the slope of $\log L/\log X$ curve as $d\log L/d\log
X=-1/\lambda$. This requires the log-likelihood curve to be concave as in
Fig.~\ref{fig:phase_chg}(a). If the log-likelihood curve is non-concave as in
Fig.~\ref{fig:phase_chg}(b), then increasing $\lambda$ from 0 to 1 will normally
take the samples from A to the neighbourhood of B where the
slope is $-1/\lambda=-1$. In order to get the samples beyond B, $\lambda$ will
need to be taken beyond 1. Doing this will take the samples around the
neighbourhood of the point of inflection C but here thermodynamic integration
sees a phase change and has to jump across, somewhere near F, in which any
practical computation exhibits hysteresis that destroys the calculation of
$\mathcal{Z}$. As will be discussed in the next section, nested sampling does 
not experience any problem with phase changes and moves steadily down in the 
prior volume $X$ regardless of whether the log-likelihood is concave or convex or 
even differentiable at all.

\section{Nested sampling}
\label{section:nested} 

\begin{figure}
 \begin{center}
    	\subfigure[]{
          \includegraphics[width=0.4\columnwidth]{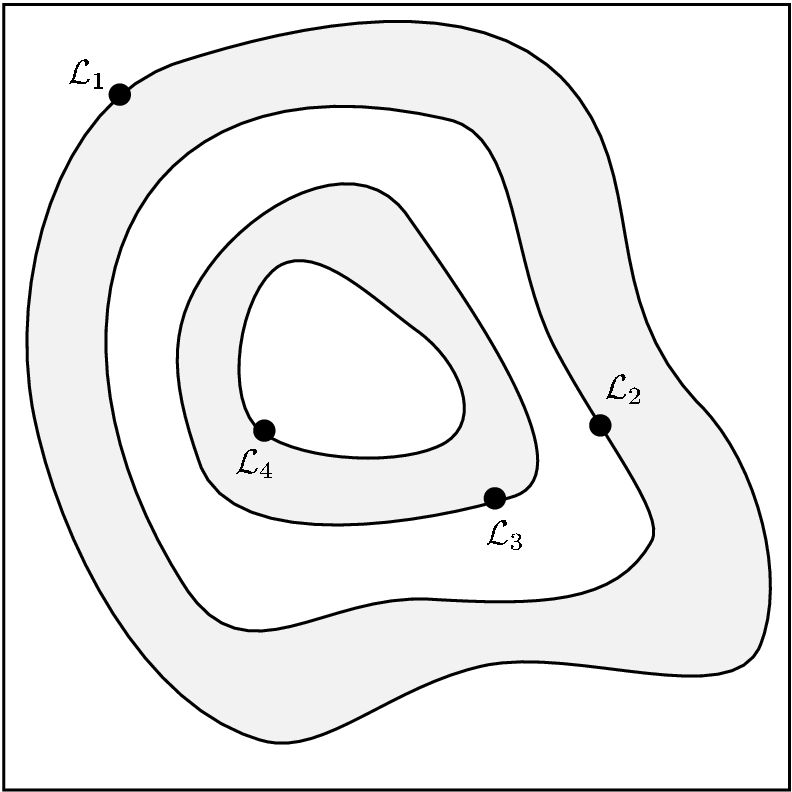}}
	  \hspace{0.3cm}
	\subfigure[]{
          \includegraphics[width=0.4\columnwidth]{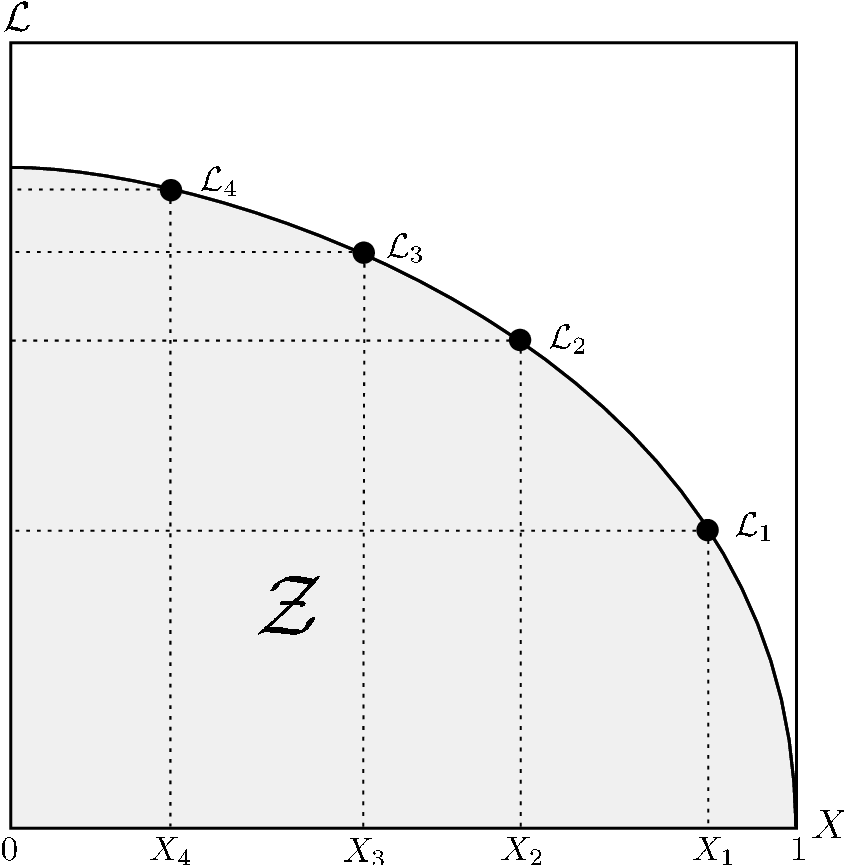}}
\caption{Cartoon illustrating (a) the posterior of a two dimensional problem; 
and (b) the transformed $L(X)$ function where the prior volumes $X_i$ are
associated with each likelihood $L_i$.}
\label{figure1}
\end{center}
\end{figure}

Nested sampling (Skilling 2004) is a Monte Carlo
technique aimed at efficient evaluation of the Bayesian
evidence, but also produces posterior inferences as a by-product. It
exploits the relation between the likelihood and prior volume to
transform the multidimensional evidence integral (\ref{eq:3}) into a
one-dimensional integral. The `prior volume' $X$ is defined by $dX =
\pi(\mathbf{\Theta})d^D \mathbf{\Theta}$, so that
\begin{equation}
X(\lambda) 
= \int_{L\left(\mathbf{\Theta}\right) > \lambda} \pi(\mathbf{\Theta}) d^D
\mathbf{\Theta},
\label{Xdef}
\end{equation}
where the integral extends over the region(s) of parameter space
contained within the iso-likelihood contour
$L(\mathbf{\Theta}) = \lambda$.  Assuming that
$L(X)$, i.e. the inverse of (\ref{Xdef}),
is a monotonically decreasing function of $X$ (which
is trivially satisfied for most posteriors), the evidence integral
(\ref{eq:3}) can then be written as
\begin{equation}
\mathcal{Z}=\int_0^1{L(X)}dX.
\label{equation:nested}
\end{equation}
Thus, if one can evaluate the likelihoods $L_{j}=L(X_{j})$, where $X_{j}$
is a sequence of decreasing values,
\begin{equation}
0<X_{M}<\cdots <X_{2}<X_{1}< X_0=1,\label{eq:5}
\end{equation}
as shown schematically in Fig.~\ref{figure1}, the evidence can be
approximated numerically using standard quadrature methods as a weighted sum
\begin{equation}
\mathcal{Z}={\textstyle {\displaystyle
\sum_{i=1}^{M}}L_{i}w_{i}}.\label{eq:6}
\end{equation} 
In the following we will use the simple trapezium rule, for which
the weights are given by $w_i=\frac{1}{2}(X_{i-1}-X_{i+1})$.
An example of a posterior in two dimensions and its associated
function $L(X)$ is shown in Fig.~\ref{figure1}.

\subsection{Evidence evaluation}
\label{nested:evidence}

The nested sampling algorithm performs the summation (\ref{eq:6}) as follows.
To begin, the iteration counter is set to $i=0$ and $N$ `live' (or
`active') samples
are drawn from the full prior $\pi(\mathbf{\Theta})$ (which is often
simply the uniform distribution over the prior range), so the initial
prior volume is $X_0=1$. The
samples are then sorted in order of their likelihood and the smallest
(with likelihood $L_0$) is removed from the live set and replaced by a
point drawn from the prior subject to the constraint that the point
has a likelihood $L>L_0$.  The corresponding prior volume contained within
this iso-likelihood contour will be a random variable given by $X_1 = t_1
X_0$, where $t_1$ follows the distribution $\Pr(t) = Nt^{N-1}$
(i.e. the probability distribution for the largest of $N$ samples
drawn uniformly from the interval $[0,1]$). At each subsequent
iteration $i$, the discarding of the lowest
likelihood point $L_i$ in the live set, the drawing of a replacement 
with $L > L_i$ and the reduction of the corresponding prior volume
$X_i=t_iX_{i-1}$ are repeated, until the entire prior volume has been
traversed. The algorithm thus travels through nested
shells of likelihood as the prior volume is reduced.

The mean and standard deviation of $\ln t$, which dominates the
geometrical exploration, are:
\begin{equation}
E[\ln t]=-\frac{1}{N},\qquad \sigma[\ln t]=\frac{1}{N}.\label{eq:8}
\end{equation}
Since each value of $\ln t$ is independent, after $i$ iterations the 
prior volume will shrink down such that
$\ln X_{i}\approx-(i\pm\sqrt{i})/N$. Thus, one takes
$X_i = \exp(-i/N)$.

\subsection{Stopping criterion}
\label{nested:stopping}

The nested sampling algorithm should be terminated on determining the
evidence to some specified precision.  One way would be to proceed
until the evidence estimated at each replacement changes by less than
a specified tolerance. This could, however, underestimate the evidence
in (for example) cases where the posterior contains any narrow peaks
close to its maximum. \citet{Skilling} provides an adequate and robust
condition by determining an upper limit on the evidence that can be
determined from the remaining set of current active points.
By selecting the maximum-likelihood $L_{\rm max}$ in the set
of active points, one can safely assume that the largest evidence
contribution that can be made by the remaining portion of the
posterior is $\Delta{\mathcal{Z}}_{\rm i} = L_{\rm max}X_{\rm i}$, i.e. the
product of the remaining prior volume and maximum likelihood value. We choose
to stop when this quantity would no longer change the final evidence
estimate by some user-defined value (we use 0.1 in log-evidence).

\begin{figure*}
 \begin{center}
    	\subfigure[]{
          \includegraphics[width=0.39\columnwidth]{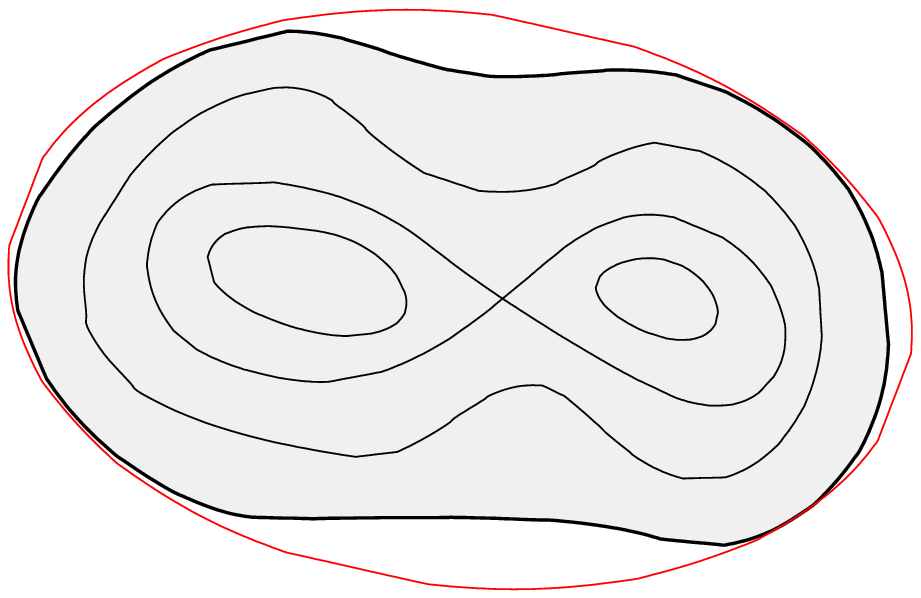}}
	\setcounter{subfigure}{1}
	\subfigure[]{
          \includegraphics[width=0.39\columnwidth]{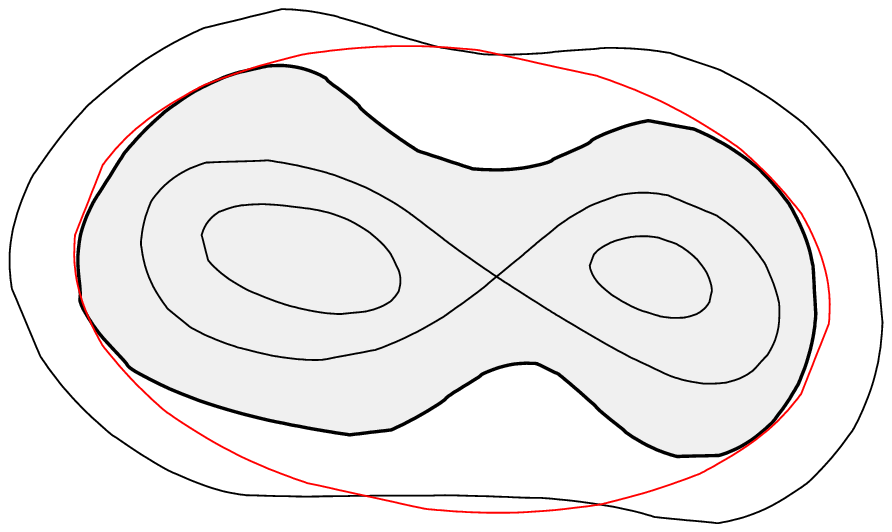}}
	\setcounter{subfigure}{2}
	\subfigure[]{
          \includegraphics[width=0.39\columnwidth]{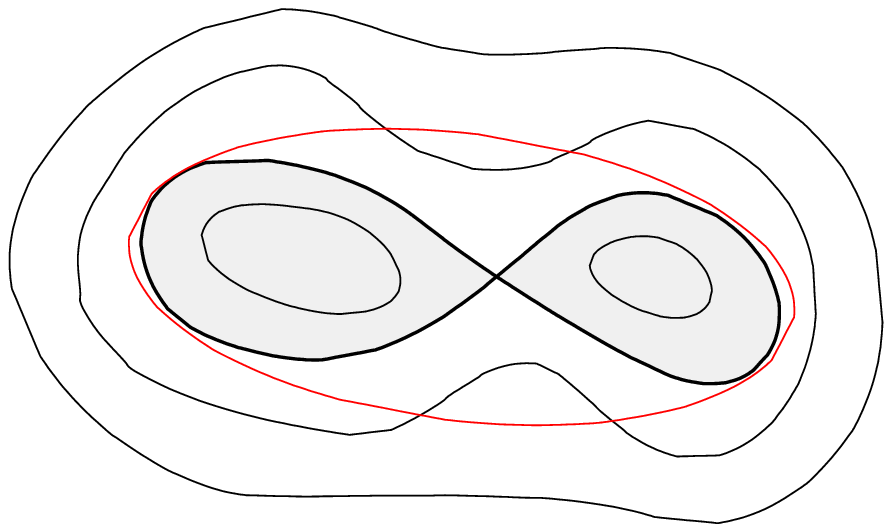}}
	\setcounter{subfigure}{3}
	\subfigure[]{
          \includegraphics[width=0.39\columnwidth]{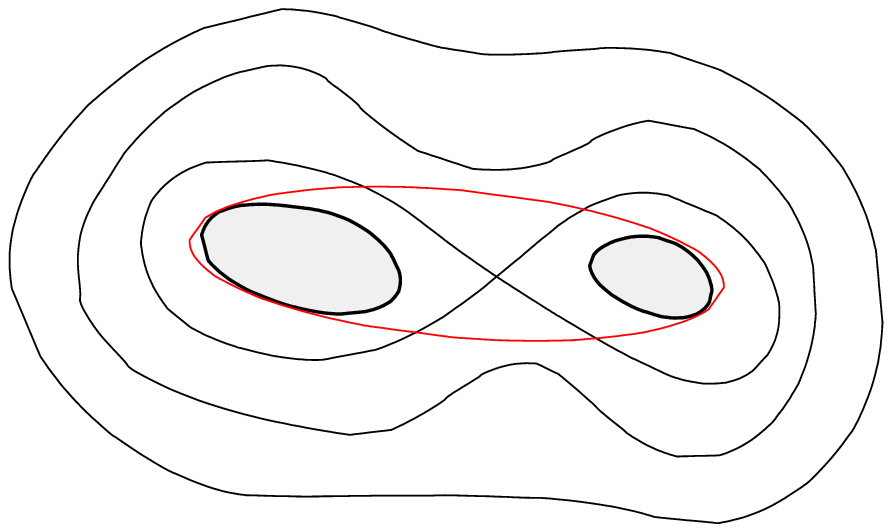}}
	\setcounter{subfigure}{4} 
	 \subfigure[]{
          \includegraphics[width=0.39\columnwidth]{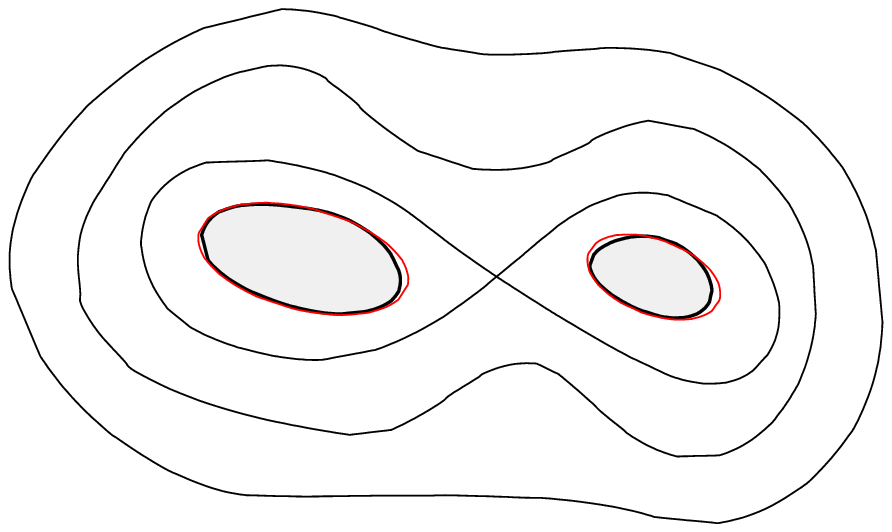}}
	  \setcounter{subfigure}{5} 
          \caption{Cartoon of ellipsoidal nested sampling from a
simple bimodal distribution. In (a) we see that the ellipsoid
represents a good bound to the active region. In (b)-(d), as we nest
inward we can see that the acceptance rate will rapidly decrease as
the bound steadily worsens. Figure (e) illustrates the increase in
efficiency obtained by sampling from each clustered region separately.}
\label{figure2}
\end{center}
\end{figure*}

\subsection{Posterior inferences}
\label{nested:posterior}

Once the evidence $\mathcal{Z}$ is found, posterior inferences can be easily
generated using the full sequence of discarded points from the
nested sampling process, i.e. the points with the
lowest likelihood value at each iteration $i$ of the algorithm. Each such point
is simply assigned the weight 
\begin{equation}
p_{i}=\frac{L_{i}w_{i}}{\mathcal{Z}}.\label{eq:12}
\end{equation}
These samples can then be used to calculate inferences of posterior 
parameters such as means, standard deviations, covariances and so on,
or to construct marginalised posterior distributions.

\subsection{Evidence error estimation}
\label{nested:error}

If we could assign each $X_i$ value exactly then the only error in our
estimate of the evidence would be due to the discretisation of the
integral (\ref{eq:6}). Since each $t_i$ is a random variable, however,
the dominant source of uncertainty in the final $\mathcal{Z}$ value
arises from the incorrect assignment of each prior volume. Fortunately,
this uncertainty can be easily estimated.

Shaw et al. made use of the knowledge of the distribution $\Pr(t_i)$
from which each $t_i$ is drawn to assess the errors in any quantities
calculated.  Given the probability of the vector
$\mathbfss{t}=(t_1,t_2,\ldots,t_M)$ as
\begin{equation}
\Pr(\mathbfss{t}) = \prod_{i=1}^M \Pr(t_i),
\end{equation}
one can write the expectation value of any quantity $F(\mathbfss{t})$ as
\begin{equation}
\langle F \rangle = \int F(\mathbfss{t})\Pr(\mathbfss{t}) d^M\mathbfss{t}.
\end{equation}
Evaluation of this integral is possible by Monte Carlo methods by
sampling a given number of vectors $\mathbfss{t}$ and finding the
average $F$. By this method one can determine the variance of the curve
in $X-L$ space, and thus the uncertainty in the evidence integral
$\int L(X)dX$. As demonstrated by Shaw et al., this eliminates the
need for any repetition of the algorithm to determine the
standard error on the evidence value; this constitutes a significant
increase in efficiency.

In our new methods presented below, however, we use a different error
estimation scheme suggested by \citet{Skilling}; this also provides an
error estimate in a single sampling but is far less computationally
expensive and proceeds as follows.  The usual behaviour of the evidence
increments $L_{i}w_{i}$ is initially to rise with iteration number
$i$, with the likelihood $L_i$ increasing faster than the weight
$w_i=\frac{1}{2}(X_{i-1}-X_{i+1})$ decreases. At some point $L$
flattens off sufficiently that the decrease in the weight dominates
the increase in likelihood, so the increment $L_iw_i$ reaches a 
maximum and then starts to drop with iteration number. Most of the
contribution to the final evidence value usually comes from the
iterations around the maximum point, which occurs in the region of
$X\approx e^{-H}$, where $H$ is the negative \emph{relative entropy},
\begin{equation}
H=\int \,\ln \left(\frac{dP}{dX}\right)\,dX\approx \sum_{i=1}^M\frac{L_{i}w_{i}}{\mathcal{Z}}
\ln\left(\frac{L_{i}}{\mathcal{Z}}\right),
\label{eq:info}
\end{equation}
where $P$ denotes the posterior. Since $\ln X_i\approx (-i\pm \sqrt{i})/N$, we expect the procedure to
take about $NH \pm \sqrt{NH}$ steps to shrink down to the bulk of the
posterior. The dominant uncertainty in ${\cal Z}$ is due to the Poisson
variability $NH \pm \sqrt{NH}$ in the number of steps to reach the
posterior bulk. Correspondingly the accumulated values $\ln X_i$ are subject
to a standard deviation uncertainty of $\sqrt{H/N}$. This uncertainty
is transmitted to the evidence $\mathcal{Z}$ through \eqref{eq:6}, so that
$\ln \mathcal{Z}$ also has standard deviation uncertainty of
$\sqrt{H/N}$. Thus, putting the results together gives
\begin{equation}
\ln\mathcal{Z}={\ln \left( \textstyle 
{\displaystyle\sum_{i=1}^{M}}L_{i}w_{i}\right)}
\pm \sqrt{\frac{H}{N}}.
\label{eq:err}
\end{equation} 

Alongside the above uncertainty, there is also the error
due to the discretisation of the integral in \eqref{eq:6}. 
Using the trapezoidal rule, this error will be
$\mathcal{\mathcal{O}}(1/M^{2})$, and hence will be negligible given
a sufficient number of iterations.  

\section{Ellipsoidal nested sampling}
\label{ellipsoidal_sampling}

The most challenging task in implementing the nested sampling
algorithm is drawing samples from the prior within the hard constraint
$L>L_i$ at each iteration $i$. Employing a naive approach that draws
blindly from the prior would result in a steady decrease in the
acceptance rate of new samples with decreasing prior volume (and
increasing likelihood).

\subsection{Single ellipsoid sampling}

Ellipsoidal sampling (\citet{Mukherjee}) partially overcomes the
above problem by approximating the iso-likelihood contour of the point
to be replaced by an $D$-dimensional ellipsoid determined from the
covariance matrix of the current set of live points. This ellipsoid is
then enlarged by some factor $f$ to account for the iso-likelihood
contour not being exactly ellipsoidal. New points are then selected
from the prior within this (enlarged) ellipsoidal bound until one is
obtained that has a likelihood exceeding that of the discarded
lowest-likelihood point. In the limit that the ellipsoid coincides
with the true iso-likelihood contour, the acceptance rate tends to
unity. An elegant method for drawing uniform samples from an
$D$-dimensional ellipsoid is given by \citet{Shaw}. and is easily
extended to non-uniform priors.

\subsection{Recursive clustering}

Ellipsoidal nested sampling as described above is efficient for simple
unimodal posterior distributions, but is not well suited to multimodal
distributions. The problem is illustrated in Fig.~\ref{figure2}, in
which one sees that the sampling efficiency from a single ellipsoid
drops rapidly as the posterior value increases (particularly in higher
dimensions). As advocated by Shaw
et al., and illustrated in the final panel of the figure, the
efficiency can be substantially improved by identifying distinct
\emph{clusters} of live points that are well separated and
constructing an individual ellipsoid for each cluster. The linear nature
of the evidence means it is valid to consider each cluster
individually and sum the contributions provided one correctly assigns
the prior volumes to each distinct region. Since the collection of $N$
active points is distributed evenly across the prior one can safely
assume that the number of points within each clustered region is
proportional to the prior volume contained therein.

\citet{Shaw} identify clusters recursively. Initially, at each
iteration $i$ of the nested sampling algorithm, $k$-means clustering
(see e.g. \citet{MacKay}) with $k=2$ is applied to the live set of points
to partition them into two clusters and an (enlarged) ellipsoid is constructed
for each one. This division of the live set will only be accepted if
two further conditions are met: (i) the total volume of the two ellipsoids 
is less than some fraction of the original pre-clustering
ellipsoid and (ii) clusters are sufficiently separated by some
distance to avoid overlapping regions. If these conditions are satisfied
clustering will occur and the number of live points in each cluster
are topped-up to $N$ by sampling from the prior inside the
corresponding ellipsoid, subject to the hard constraint $L>L_i$.
The algorithm then searches independently within each cluster attempting to
divide it further. This process continues recursively until the
stopping criterion is met. Shaw et al. also show how the error
estimation procedure can be modified to accommodate clustering by 
finding the probability distribution of the volume fraction in each cluster.

\section{Improved ellipsoidal sampling methods}
\label{improvements}

In this section, we present two new methods for ellipsoidal nested
sampling that improve significantly in terms of sampling efficiency
and robustness on the existing techniques outlined above, in
particular for multimodal distributions and those with pronounced
degeneracies.

\subsection{General improvements}
\label{improvements:general}

We begin by noting several general improvements that are employed by
 one or other of our new methods.

\subsubsection{Identification of clusters}
\label{improvements:identify_cluster}
In both methods, we wish to identify isolated modes of the posterior
distribution without prior knowledge of their number. The only
information we have is the current live point set. Rather than using
$k$-means clustering with $k=2$ to partition the points into just two
clusters at each iteration, we instead attempt to infer the
appropriate number of clusters from the point set. After experimenting
with several clustering algorithms to partition the points into the
optimal number of clusters, we found X-means \citep{Pelleg}, G-means
\citep{Hamerly} and PG-means \citep{Feng} to be the most
promising. X-means partitions the points into the number of clusters
that optimizes the Bayesian Information Criteria (BIC) measure. The
G-means algorithm is based on a statistical test for the hypothesis
that a subset of data follows a Gaussian distribution and runs
$k$-means with increasing $k$ in a hierarchical fashion until the test
accepts the hypothesis that the data assigned to each $k$-means centre
are Gaussian. PG-means is an extension of G-means that is able to
learn the number of clusters in the classical Gaussian mixture model
without using $k$-means. We found PG-means to outperform both X-means
and G-means, especially in higher dimensions and if there are cluster
intersections, but the method requires Monte Carlo simulations at each
iteration to calculate the critical values of the Kolmogorov--Smirnov
test it uses to check for Gaussianity. As a result, PG-means is
considerably more computationally expensive than both X-means and
G-means, and this computational cost quickly becomes
prohibitive. Comparing X-means and G-means, we found the former to
produce more consistent results, particularly in higher
dimensions. Since we have to cluster the live points at each iteration
of the nested sampling process, we thus chose to use the X-means
clustering algorithm. This method performs well overall, but does
suffers from some occasional problems that can result in the number of
clusters identified being more or less than the actual number. We
discuss these problems in the context of both our implementations in
sections~\ref{method:simultaneous} and \ref{method:clustered} but
conclude they do not adversely affect out methods.  Ideally, we
require a fast and robust clustering algorithm that always produces
reliable results, particularly in high dimensions. If such a method
became available, it could easily be substituted for X-means in either
of our sampling techniques described below.

\subsubsection{Dynamic enlargement factor}
\label{improvements:def}
Once an ellipsoid has been constructed for each identified cluster
such that it (just) encloses all the corresponding live points, it is
enlarged by some factor $f$, as discussed in
Sec.~\ref{ellipsoidal_sampling}. It is worth remembering that the corresponding
increase in volume is $(1+f)^D$, where $D$ is the dimension of the parameter
space. The factor $f$ does not, however,
have to remain constant. Indeed, as the nested sampling algorithm
moves into higher likelihood regions (with decreasing prior volume), the
enlargement factor $f$ by which an ellipsoid is expanded can be made
progressively smaller. This holds since the ellipsoidal approximation
to the iso-likelihood contour obtained from the $N$ live points
becomes increasingly accurate with decreasing prior volume.

Also, when more than one ellipsoid is constructed at some iteration, the
ellipsoids with fewer points require a higher enlargement factor
than those with a larger number of points. This is due to the error
introduced in the evaluation of the eigenvalues from the covariance
matrix calculated from a limited sample size. The standard deviation
uncertainty in the eigenvalues is given by \citet{Girshick} as
follows:
\begin{equation}
\sigma(\hat{\lambda}_{j})\approx\lambda_{j}\sqrt{2/n},
\label{eq:evs}
\end{equation}
where $\lambda_{j}$ denotes the $j$th eigenvalue and $n$ is the number
of points used in the calculation of the covariance matrix.

The above considerations lead us to set the enlargement factor for the
$k$th ellipsoid at iteration $i$ as $f_{i,k} =
f_{0}X_{i}^\alpha\sqrt{N/n_k}$ where $N$ is the total number of live
points, $f_0$ is the initial user--defined enlargement factor (defining
the percentage by which each axis of an ellipsoid enclosing $N$ points, is 
enlarged), $X_i$ is the prior volume at the $i$th iteration, $n_k$ is the 
number of points in the $k^{th}$ cluster, and $\alpha$ is a value between $0$ 
and $1$ that defines the rate at which the enlargement factor decreases with
decreasing prior volume.

\subsubsection{Detection of overlapping ellipsoids}
\label{improvements:detect_overlap}
In some parts of our sampling methods, it is important to have a very
fast method to determine whether two ellipsoids intersect, as this
operation is performed many times at each iteration.  Rather than
applying the heuristic criteria used by Shaw et al., we instead employ
an exact algorithm proposed by \citet{Alfano} which involves the
calculation of eigenvalues and eigenvectors of the covariance matrix
of the points in each ellipsoid. Since we have already calculated
these quantities in constructing the ellipsoids, we can rapidly
determine if two ellipsoids intersect at very little extra
computational cost.

\subsubsection{Sampling from overlapping ellipsoids}
\label{improvements:sample_overlap}

\begin{figure}
\begin{center}
\includegraphics[width=0.7\columnwidth]{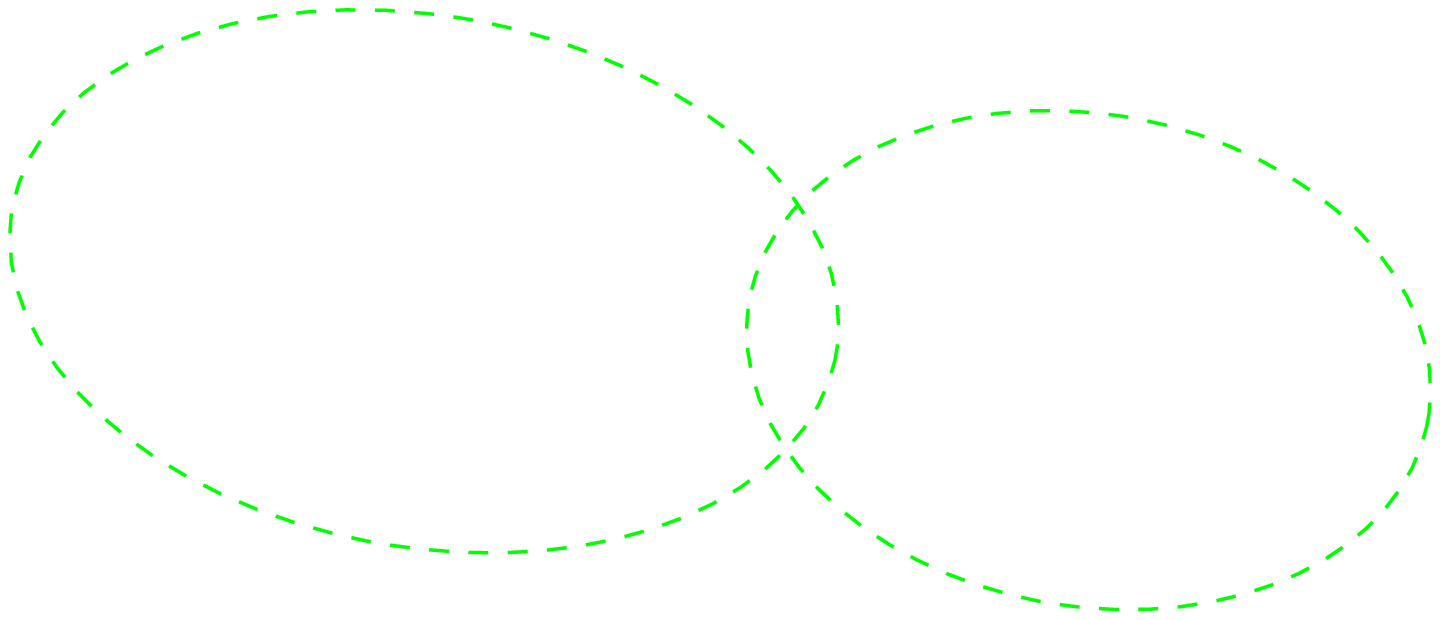}
\caption{If the ellipsoids corresponding to different modes are
overlapping then sampling from one ellipsoid, enclosing all the
points, can be quite inefficient. Multiple overlapping ellipsoids
present a better approximation to the iso-likelihood contour of a
multimodal distribution.}
\label{overlap}
\end{center}
\end{figure}

As illustrated earlier in Fig.~\ref{figure2}, for a multimodal
distribution multiple ellipsoids represent a much better approximation
to the iso-likelihood contour than a single ellipsoid containing all
the live points. At likelihood levels around which modes
separate, X-means will often partition the point set into a
number of distinct clusters, but the (enlarged) ellipsoids enclosing
distinct identified clusters will tend to overlap (see
Fig.~\ref{overlap}) and the partitioning will be discarded. At some
sufficiently higher likelihood level, the corresponding ellipsoids
will usually no longer overlap, but it is wasteful to wait for this to
occur. Hence, in both of our new sampling methods described below it
will prove extremely useful to be able to sample consistently from
ellipsoids that may be overlapping, without biassing the resultant
evidence value or posterior inferences.

Suppose at iteration $i$ of the nested
sampling algorithm, a set of live points is partitioned into $K$
clusters by X-means, with the $k^{th}$ cluster having $n_{k}$
points. Using the covariance matrices of each set of points, each
cluster then is enclosed in an ellipsoid which is then expanded using
an enlargement factor $f_{i,k}$. The volume $V_{k}$ of each resulting
ellipsoid is then found and one ellipsoid is chosen with probability
$p_{k}$ equal to its volume fraction:
\begin{equation}
p_{k}=V_{k}/V_{\rm tot},
\label{eq:13}
\end{equation}
where $V_{\rm tot} = \sum_{k=1}^{K}V_{k}$. Samples are then drawn
from the chosen ellipsoid until a sample is found for which the hard
constraint $L>L_i$ is satisfied, where $L_i$ is the
lowest-likelihood value among all the live points under
consideration. There is, of course, a possibility that the chosen
ellipsoid overlaps with one or more other ellipsoids. In order to
take an account of this possibility, we find the number of ellipsoids,
$n_e$, in which the sample lies and only accept the sample with
probability $1/n_e$. This provides a consistent sampling procedure in
all cases.

\subsection{Method 1: simultaneous ellipsoidal sampling}
\label{method:simultaneous}

This method is built in large part around the above technique for
sampling consistently from potentially overlapping ellipsoids. At
each iteration $i$ of the nested sampling algorithm, the method
proceeds as follows. The full set of $N$ live points is partitioned
using X-means, which returns $K$ clusters with $n_1,n_2,\ldots,n_K$
points respectively. For each cluster, the covariance matrix of the
points is calculated and used to construct an ellipsoid that just
encloses all the points; each ellipsoid is then expanded by the
enlargement factor $f_{i,k}$ (which can depend on iteration number $i$ 
as well as the number of points in the $k$th ellipsoid, as outlined above). 
This results in a set of $K$ ellipsoids $e_1, e_2, \ldots, e_K$ at 
each iteration, which we refer to as sibling ellipsoids. The 
lowest-likelihood point (with likelihood $L_i$) from the full set of 
$N$ live points is then discarded and replaced by a new point drawn 
from the set of sibling ellipsoids, correctly taking into account any 
overlaps.

It is worth noting that at early iterations of the nested sampling
process, X-means usually identifies only $K=1$ cluster and the
corresponding (enlarged) ellipsoid completely encloses the prior
range, in which case sampling is performed from the prior range
instead. Beyond this minor inconvenience, it is important to recognise
that any drawbacks of the X-means clustering method have little impact
on the accuracy of the calculated evidence or posterior inferences.
We use X-means only to limit the remaining prior space from which to
sample, in order to increase efficiency.  If X-means returns greater
or fewer than the desired number of clusters, one would still sample
uniformly from the remaining prior space since the union of the
corresponding (enlarged) ellipsoids would still enclose all the
remaining prior volume. Hence, the evidence calculated and posterior
inferences would remain accurate to within the uncertainties discussed
in Sec.~\ref{nested:error}.

\subsection{Method 2: clustered ellipsoidal sampling}
\label{method:clustered}

This method is closer in spirit to the recursive clustering technique
advocated by Shaw et al. At the $i$th iteration of the nested sampling
algorithm, the method proceeds as follows. The full set of $N$ live
points is again partitioned using X-means to obtain $K$ clusters with
$n_{1},n_{2},\,...,\, n_{K}$ points respectively, and each cluster is
enclosed in an expanded ellipsoid as outlined above.  In this second
approach, however, each ellipsoid is then tested to determine if it
intersects with any of its sibling ellipsoids or any other
non-ancestor ellipsoid\footnote{A non-ancestor ellipsoid of $e_{k}$ is
any ellipsoid that was non-intersecting at an earlier iteration and
does not completely enclose $e_{k}$.}.  The nested sampling algorithm
is then continued {\em separately} for each cluster contained within a
non-intersecting ellipsoid $e_{k}$, after in each case (i) topping up
the number of points to $N$ by sampling $N-n_k$ points within $e_k$
that satisfy $L > L_i$; and (ii) setting the corresponding remaining
prior volume to $X_i^{(k)} = X_{i-1}(n_k/N)$. Finally, the remaining set
of $N_r$ points contained within the union of the intersecting
ellipsoids at iteration $i$ is topped up to $N$ using the method for
sampling from such a set of ellipsoids outlined in 
Sec.~\ref{improvements:sample_overlap}, and the associated remaining prior volume 
is set to $X_i=X_{i-1}(N_r/N)$.

As expected, in the early stages, X-means again usually identifies
only $K=1$ cluster and this is dealt with as in Method 1.  Once
again, the drawbacks of X-means do not have much impact on the
accuracy of the global evidence determination. If X-means finds fewer
clusters than the true number of modes, then some clusters correspond
to more than one mode and will have an enclosing ellipsoid larger than
it would if X-means had done a perfect job; this increases the chances of
the ellipsoid intersecting with some of its sibling or non-ancestor
ellipsoids. If this ellipsoid is non-intersecting, then it can still
split later and hence we do not lose accuracy. On the other hand, if
X-means finds more clusters than the true number of modes, it is again
likely that the corresponding enclosing ellipsoids will overlap. It is
only in the rare case where some of such ellipsoids are non-intersecting,
that the possibility exists for missing part of the true prior volume.
Our use of an enlargement factor strongly mitigates against this
occurring. Indeed, we have not observed such behaviour in any of our
numerical tests.

\subsection{Evaluating `local' evidences}
\label{method:levidence}

For a multimodal posterior, it can prove useful to estimate not only
the total (global) evidence, but also the `local' evidences associated
with each mode of the distribution. There is inevitably some
arbitrariness in defining these quantities, since modes of the
posterior necessarily sit on top of some general `background' in the
probability distribution. Moreover, modes lying close to one another
in the parameter space may only `separate out' at relatively high
likelihood levels. Nonetheless, for well-defined, isolated modes, a
reasonable estimate of the posterior volume that each contains (and
hence the local evidence) can be defined and estimated. Once the
nested sampling algorithm has progressed to a likelihood level such
that (at least locally) the `footprint' of the mode is well-defined,
one needs to identify at each subsequent iteration those points in the
live set belonging to that mode.  The practical means of performing this
identification and evaluating the local evidence for each mode differs
between our two sampling methods.

\subsubsection{Method 1} 
The key feature of this method is that at each
iteration the full live set of $N$ points is evolved by replacing the
lowest likelihood point with one drawn (consistently) from the
complete set of (potentially overlapping) ellipsoids.  Thus, once a
likelihood level is reached such that the footprint of some mode is
well defined, to evaluate its local evidence one requires that at each
subsequent iteration the points associated with the mode are
consistently identified as a single cluster. If such an
identification were possible, at the $i$th iteration one would simply
proceeds as follows: (i) identify the cluster (contained within the
ellipsoid $e_l$) to which the point with the lowest likelihood $L_i$
value belongs; (ii) update the local prior volume of each of the
clusters as $X_{i}^{(k)}=(n_{k}/N)X_{i}$, where $n_{k}$ is the number
of points belonging to the $k$th cluster and $X_i$ is the total
remaining prior volume; (iii) increment the local evidence of the
cluster contained within $e_{l}$ by
$\frac{1}{2}L_{i}(X_{i-1}^{(l)}-X_{i+1}^{(l)})$. Unfortunately, we
have found that X-means is not capable of consistently identifying the
points associated with some mode as a single cluster. Rather, the
partitioning of the live point set into clusters can vary appreciably
from one iteration to the next. PG-means produced reasonably
consistent results, but as mentioned above is far too computationally
intensive. We are currently exploring ways to reduce the most
computationally expensive step in PG-means of calculating the critical
values for Kolmogorov--Smirnov test, but this is not yet completed.
Thus, in the absence of a fast and consistent clustering algorithm, it
is currently not possible to calculate the local evidence of each mode
with our simultaneous ellipsoidal sampling algorithm.

\subsubsection{Method 2}
The key feature of this method is that once a cluster
of points has been identified such that its (enlarged) enclosing
ellipsoid does not intersect with any of its sibling ellipsoids (or any
other non-ancestor ellipsoid), that set of points is evolved {\em
independently} of the rest (after topping up the number of points in
the cluster to $N$). This approach therefore has some natural
advantages in evaluating local evidences. There remain, however, some
problems associated with modes that are sufficiently close to one
another in the parameter space that they are only identified as
separate clusters (with non-intersecting enclosing ellipsoids) once the
algorithm has proceeded to likelihood values somewhat larger than the value
at which the modes actually separate. In such cases, the local
evidence of each mode will be underestimated. The simplest solution to
this problem would be to increment the local evidence of each cluster
even if its corresponding ellipsoid intersects with other ellipsoids,
but as mentioned above X-means cannot produce the consistent
clustering required. In this case we have the advantage of knowing the
iteration beyond which a non-intersecting ellipsoid is regarded as a
separate mode (or a collection of modes) and hence we can circumvent 
this problem by storing information (eigenvalues, eigenvectors, 
enlargement factors etc.) of all the clusters identified, as well as 
the rejected points and their likelihood values, from the last few
iterations. We then attempt to match the clusters in the current
iteration to those identified in the last few iterations, allowing for
the insertion or rejection of points from clusters during the
intervening iterations On finding a match for some cluster in a
previous iteration $i'$, we check to see which (if any) of the points
discarded between the iteration $i'$ and the current iteration $i$
were members of the cluster. For each iteration $j$ (between $i'$ and
$i$ inclusive) where this occurs, the local evidence of the cluster is
incremented by $L_{j}X_{j}$, where $L_{j}$ and $X_{j}$ are the lowest
likelihood value and the remaining prior volume corresponding to
iteration $j$. This series of operations can be performed quite
efficiently; even storing information as far as back as 15 iterations
does not increase the running time of the algorithm appreciably.
Finally, we note that if closely lying modes have
very different amplitudes, the mode(s) with low amplitude may never be
identified as being separate and will eventually be lost as the
algorithm moves to higher likelihood values.

\subsection{Dealing with degeneracies}
\label{method:degeneracies}

\begin{figure}
\begin{center}
\includegraphics[width=0.6\columnwidth]{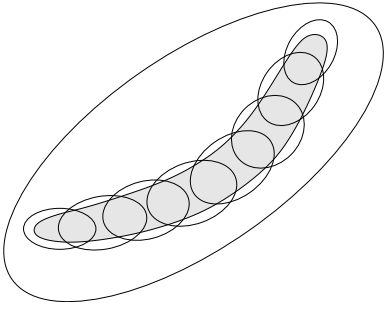}
\caption{Cartoon of the sub-clustering approach used to deal with
degeneracies. The true iso-likelihood contour contains the shaded
region. The large enclosing ellipse is typical of that constructed
using our basic method, whereas sub-clustering produces
the set of small ellipses.}
\label{fig:degen}
\end{center}
\end{figure}

As will be demonstrated in Sec.~\ref{applications}, the above methods are very
efficient and robust at sampling from multimodal distributions where
each mode is well-described at most likelihood levels by a
multivariate Gaussian. Such posteriors might be described colloquially
as resembling a `bunch of grapes' (albeit in many dimensions). In some
problems, however, some modes of the posterior might possess a
pronounced curving degeneracy so that it more closely resembles a
(multidimensional) `banana'. Such features are problematic for
all sampling methods, including our proposed ellipsoidal sampling
techniques. Fortunately, we have found that a simple modification
to our methods allows for efficient sampling even in the presence of
pronounced degeneracies.

The essence of the modification is illustrated in
Fig.~\ref{fig:degen}. Consider an isolated mode with an
iso-likelihood contour displaying a pronounced curved
degeneracy. X-means will usually identify all the live points
contained within it as belonging to a single cluster and hence the
corresponding (enlarged) ellipsoid will represent a very poor
approximation. If, however, one divides each cluster identified by
X-means into a set of {\em sub-clusters}, one can more accurately
approximate the iso-likelihood contour with many small overlapping
ellipsoids and sample from them using the method outlined in
Sec.~\ref{improvements:sample_overlap}.

To sample with maximum efficiency from a pronounced degeneracy
(particularly in higher dimensions), one would like to divide every
cluster found by X-means into as many sub-clusters as possible to allow maximum
flexibility in following the degeneracy. In
order to be able to calculate covariance matrices, however, each
sub-cluster must contain at least $(D+1)$ points, where $D$ is the
dimensionality of the parameter space. This in turn sets an upper limit
on the number of sub-clusters.

Sub-clustering is performed through an incremental $k$-means algorithm
with $k=2$. The process starts with all the points assigned to the
original cluster. At iteration $i$ of the algorithm, a point is picked
at random from the sub-cluster $c_j$ that contains the most
points. This point is then set as the centroid, $m_{i+1}$, of a new cluster
$c_{i+1}$. All those points in any of the other sub-clusters that are
closer to $m_{i+1}$ than the centroid of their own sub-cluster, and whose
sub-cluster has more than $(D+1)$ points are then assigned to
$c_{i+1}$ and $m_{i+1}$ is updated. All the points not belonging to
$c_{i+1}$ are again checked with the updated $m_{i+1}$ until no new
point is assigned to $c_{i+1}$. At the end of the iteration $i$, if
$c_{i+1}$ has less than $(D+1)$ points then the points in $c_j$ that
are closest to $m_{i+1}$ are assigned to $c_{i+1}$ until $c_{i+1}$ has
$(D+1)$ points. In the case that $c_j$ has fewer than $2(D+1)$ points,
then points are assigned from $c_{i+1}$ to $c_j$. The algorithm stops
when, at the start of an iteration, the sub-cluster with most points
has fewer than $2(D+1)$ members, since that would result in a new
sub-cluster with fewer than $2(D+1)$ points. This process can result
in quite a few sub-clusters with more than $2(D+1)$ but less than
$2(D+1)$ points and hence there is a possibility for even more
sub-clusters to be formed. This is achieved by finding the sub-cluster
$c_l$ closest to the cluster, $c_k$. If the sum of points in 
$c_l$ and $c_k$ is greater than or equal to $3(D+1)$, an additional 
sub-cluster is created out of them.

Finally, we further reduce the possibility that the union of the
ellipsoids corresponding to different sub-clusters might not enclose
the entire remaining prior volume as follows. For each sub-cluster
$c_k$, we find the one point in each of the $n$ nearest sub-clusters
that is closest to the centroid of $c_k$. Each such point is then assigned
to $c_k$ {\em and} its original sub-cluster, i.e. it is `shared'
between the two sub-clusters. In this way all the sub-clusters and
their corresponding ellipsoids are expanded, jointly enclosing the
whole of the remaining prior volume. In our numerical simulations, we
found setting $n=5$ performs well.

\section{Metropolis Nested Sampling}
\label{metronest}

An alternative method for drawing samples from the prior within the
hard constraint $L>L_i$ where $L_i$ is the lowest likelihood value at
iteration $i$, is the standard Metropolis algorithm (see e.g. \citet{MacKay}) 
as suggested in \citet{Sivia}. In this approach, at each
iteration, one of the live points, $\mathbf{\Theta}$, is picked at
random and a new trial point, $\mathbf{\Theta}^\prime$,
is generated using a symmetric proposal distribution
$Q(\mathbf{\Theta}^\prime,\mathbf{\Theta})$. The trial point
$\mathbf{\Theta}^\prime$ is then accepted with probability
\begin{equation}
\alpha = 
\begin{cases}
1 & 
\mbox{if $\pi(\mathbf{\Theta}^\prime) > \pi(\mathbf{\Theta})$ and
$L(\mathbf{\Theta}^\prime) > L_i$} \\
\pi(\mathbf{\Theta}^\prime)/\pi(\mathbf{\Theta}) & 
\mbox{if $\pi(\mathbf{\Theta}^\prime) \leq \pi(\mathbf{\Theta})$ and
$L(\mathbf{\Theta}^\prime) > L_i$} \\
0 & \mbox{otherwise}
\end{cases}
\label{metro}
\end{equation}
A symmetric Gaussian distribution is often used as the proposal
distribution. The dispersion $\sigma$ of this Gaussian should be
sufficiently large compared to the size of the region satisfying $L >
L_i$ that the chain is reasonably mobile, but without being so large
that the likelihood constraint stops nearly all proposed moves. Since
an independent sample is required, $n_{\rm step}$ steps are taken by
the Metropolis algorithm so that the chain diffuses far away from the
starting position $\mathbf{\Theta}$ and the memory of it is lost.  In
principle, one could calculate convergence statistics to determine at
which point the chain is sampling from the target distribution.
\citet{Sivia} propose, however, that one should instead simply take 
$n_{\rm step} \approx 20$ steps in all cases. The 
appropriate value of $\sigma$ tends to diminish as the nested
algorithm moves towards higher likelihood regions and decreasing
prior mass. Hence, the value of $\sigma$ is updated at the end of
each nested sampling iteration, so that the acceptance rate is around 
50\%, as follows:
\begin{equation}
\sigma \rightarrow \left\{ \begin{array}{ll}
         \sigma e^{1/N_{\rm a}} & \mbox{if $N_{\rm a} > N_{\rm r}$}\\
         \sigma e^{-1/N_{\rm r}} & \mbox{if $N_{\rm a} \le N_{\rm r}$}
\end{array} \right.,
\end{equation}
where $N_{\rm a}$ and $N_{\rm r}$ are the numbers of accepted and
rejected samples in the latest Metropolis sampling phase.

In principle, this approach can be used quite generally and does not
require any clustering of the live points or construction of
ellipsoidal bounds. In order to facilitate the evaluation of `local'
evidences, however, we combine this approach with the clustering
process performed in Method 2 above to produce a hybrid algorithm,
which we describe below. Moreover, as we show in Section~\ref{Ex1},
this hybrid approach is significantly more efficient in sampling from
multimodal posteriors than using just the Metropolis algorithm without
clustering.

At each iteration of the nested sampling process, the set of live
points is partitioned into clusters, (enlarged) enclosing ellipsoids
are constructed, and overlap detection is performed precisely in the 
clustered ellipsoidal method. Once again,
the nested sampling algorithm is then continued {\em separately} for
each cluster contained within a non-intersecting ellipsoid $e_{k}$. 
This proceeds by (i) topping up the number of
points in each cluster to $N$ by sampling $N-n_k$ points that satisfy
$L > L_i$ using the Metropolis method described above, and (ii)
setting the corresponding remaining prior mass to $X_i^{(k)} =
X_{i-1}(n_k/N)$.  Prior to topping up a cluster
in step (i), a `mini' burn-in is performed during which the width
$\sigma_k$ of the proposal distribution is adjusted as described above; 
the width $\sigma_k$ is then kept constant during the topping-up step. 

During the sampling the starting point $\mathbf{\Theta}$ for the random walk 
is chosen by picking one of 
the ellipsoids with probability $p_{k}$ equal to its volume fraction:
\begin{equation}
p_{k}=V_{k}/V_{\rm tot},
\label{eq:vol}
\end{equation}
where $V_{k}$ is the volume occupied by the ellipsoid $e_k$ and 
$V_{\rm tot} = \sum_{k=1}^{K}V_{k}$, and then picking randomly from the points 
lying inside the chosen ellipsoid. This is done so that the number of points 
inside the modes is proportional to the prior volume occupied by those modes. 
We also supplement the condition (\ref{metro}) for a trial point to be accepted 
by the requirement that it must not lie inside any of the non-ancestor 
ellipsoids in order to avoid over-sampling any region of the prior space. 
Moreover, in step (i) if any sample accepted during the topping-up step lies 
outside its corresponding (expanded) ellipsoid, then that ellipsoid is dropped 
from the list of those to be explored as an isolated likelihood region in the 
current iteration since that would mean that the region has not truly separated 
from the rest of the prior space.

Metropolis nested sampling can be quite efficient in
higher-dimensional problems as compared with the ellipsoidal sampling
methods since, in such cases, even a small region of an ellipsoid
lying outide the true iso-likelihood contour would occupy a large
volume and hence result in a large drop in efficiency. Metropolis nested
sampling method does not suffer from this curse of dimensionality as
it only uses the ellipsoids to separate the isolated likelihood
regions and consequently the efficiency remains approximately constant
at $\sim 1/n_{\rm step}$, which is $5$ per cent in our case.  This will
be illustrated in the next section in which  Metropolis nested
sampling is denoted as Method 3.

\section{Applications}
\label{applications}

In this section we apply the three new algorithms discussed in the
previous sections to two toy problems to demonstrate that
they indeed calculate the Bayesian evidence and make posterior
inferences accurately and efficiently.

\subsection{Toy model 1}
\label{Ex1}

For our first example, we consider the problem investigated by
\citet{Shaw} as their Toy Model II, which has a posterior of known
functional form so that an analytical evidence is available to compare
with those found by our nested sampling algorithms. The
two-dimensional posterior consists of the sum of 5 Gaussian peaks of
varying width, $\sigma_{k}$, and amplitude, $A_{k}$, placed randomly
within the unit circle in the $xy$-plane. The parameter values
defining the Gaussians are listed in Table~\ref{tab:Ex1_1}, leading to an
analytical total log-evidence $\ln\mathcal{Z} = -5.271$.  The
analytical `local' log-evidence associated with each of the 5 Gaussian
peaks is also shown in the table.
\begin{figure}
\begin{center}
\includegraphics[width=0.9\columnwidth]{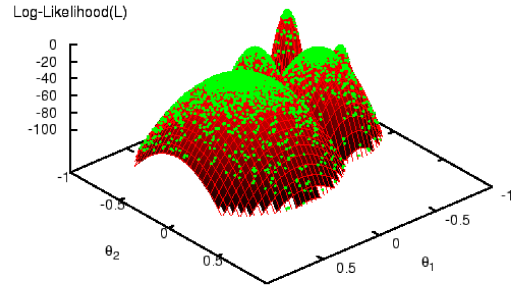}
\caption{Toy Model 1a: a two-dimensional posterior consisting of the
  sum of 5 Gaussian peaks of varying width and height placed randomly
  in the unit circle in the $xy$-plane. The dots denote the set of live 
  points at each successive likelihood level in the nested sampling 
  algorithm using Method 1 (simultaneous ellipsoidal sampling).}
\label{fig:Ex1_sns}
\end{center}
\end{figure}
\begin{figure}
\begin{center}
\includegraphics[width=0.9\columnwidth]{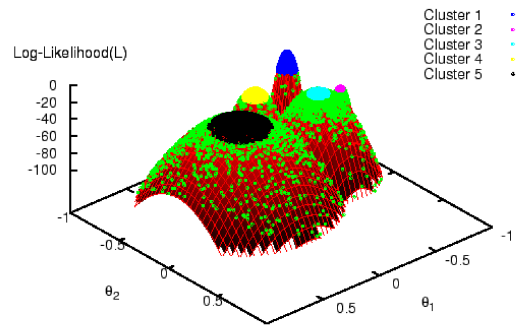}
\caption{As in Fig.~\ref{fig:Ex1_sns}, but using Method 2 (clustered
ellipsoidal sampling). The different colours denote points assigned to
isolated clusters as the algorithm progresses.}
\label{fig:Ex1_ces}
\end{center}
\end{figure}

\begin{table}
\begin{center}
\begin{tabular}{crrrrc}
\hline
Peak & $X$  & $Y$ & $A$ & $\sigma$ & Local $\ln\mathcal{Z}$ \\
\hline
1 & $-0.400$ & $-0.400$ & $0.500$ & $0.010$ & $-9.210$ \\
2 & $-0.350$ & $ 0.200$ & $1.000$ & $0.010$ & $-8.517$ \\
3 & $-0.200$ & $ 0.150$ & $0.800$ & $0.030$ & $-6.543$ \\
4 & $ 0.100$ & $-0.150$ & $0.500$ & $0.020$ & $-7.824$ \\
5 & $ 0.450$ & $ 0.100$ & $0.600$ & $0.050$ & $-5.809$ \\
\hline
\end{tabular}
\caption{The parameters $X_{k}$, $Y_{k}$, $A_{k}$, $\sigma_{k}$
defining the 5 Gaussians in Fig.~\ref{fig:Ex1_sns}. The log-volume
(or local log-evidence) of each Gaussian is also shown.}
\label{tab:Ex1_1}
\end{center}
\end{table}

\begin{table}
\begin{center}
\begin{tabular}{ccccc}
\hline
Toy model 1a & Method 1  & Method 2 & Method 3 &   Shaw et al.\\
\hline
$\ln\mathcal{Z}$&$-5.247$&$-5.178$&$-5.358$&$-5.296$\\
Error           &0.110&0.112&0.115&0.084\\
$N_{\rm like}$  &39,911&12,569&161,202&101,699\\
\hline
\end{tabular}
\caption{The calculated global log-evidence, its uncertainty and the number of
likelihood evaluations required in analysing Toy model 1a using Method 1
(simultaneous nested sampling), Method 2 (clustered
ellipsoidal sampling) and the recursive clustering method described 
by \citet{Shaw}. The values correspond to a single run of each algorithm.
The analytical global log-evidence is $-5.271$.}
\label{tab:Ex1_res}
\end{center}
\end{table}

The results of applying Method 1 (simultaneous ellipsoidal sampling),
Method 2 (clustered ellipsoidal sampling) to this problem are
illustrated in Figs~\ref{fig:Ex1_sns} and \ref{fig:Ex1_ces}
respectively; a very similar plot to Fig.~\ref{fig:Ex1_ces} is
obtained for Method 3 (Metropolis nested sampling).  For all three
methods, we used $N=300$ live points, switched off the sub-clustering
modification (for methods 1 and 2) outlined in
Sec.~\ref{method:degeneracies}, and assumed a flat prior within the
unit circle for the parameters $X$ and $Y$ in this two-dimensional
problem. In each figure, the dots denote the set of live points at
each successive likelihood level in the nested sampling algorithm. For
methods 2 and 3, the different colours denote points assigned to
isolated clusters as the algorithm progresses. We see that all three
algorithms sample effectively from all the peaks, even correctly
isolating the narrow Gaussian peak (cluster 2) superposed on the broad
Gaussian mode (cluster 3).

The global log-evidence values, their uncertainties and the number of
likelihood evaluations required for each method are shown in
Table~\ref{tab:Ex1_res}. Methods 1, 2 and 3, all produce evidence
values that are accurate to within the estimated uncertainties.  Also,
listed in the table are the corresponding quantities obtained by \citet{Shaw}, 
which are clearly consistent. Of particular interest,
is the number of likelihood evaluations required to produce these
evidence estimates. Methods 1 and 2 made around 40,000 and 10,000
likelihood evaluations respectively, whereas the Shaw et al. method
required more than 3 times this number (in all cases just one run of
the algorithm was performed, since multiple runs are not required to
estimate the uncertainty in the evidence). Method 3 required about
170,000 likelihood evaluations since its efficiency remains constant
at around 5\%. It should be remembered that Shaw et al. showed that
using thermodynamic integration, and performing 10 separate runs to
estimate the error in the evidence, required $\sim 3.6 \times 10^6$
likelihood evaluations to reach the same accuracy. As an aside, we
also investigated a `vanilla' version of the Metropolis nested
sampling approach, in which no clustering was performed. In this case,
over 570,000 likelihood evaluations were required to estimate the
evidence to the same accuracy. This drop in efficieny relative to
Method 3 resulted from having to sample inside different modes using a
proposal distribution with the same width $\sigma$ in every case. This
leads to a high rejection rate inside narrow modes and random walk
behaviour in the wider modes. In higher dimensions this effect will be
exacerbated. Consequently, the clustering process seems crucial for
sampling efficiently from multimodal distributions of different sizes
using Metropolis nested sampling.

Using methods 2 (clustered ellipsoidal sampling) and 3 (Metropolis
sampling) it is possible to calculate the `local' evidence and
make posterior inferences for each peak separately. For Method 2,
the mean values
inferred for the parameters $X$ and $Y$ and the local evidences thus
obtained are listed in Table~\ref{tab:Ex1_ces}, and clearly compare
well with the true values given in Table~\ref{tab:Ex1_1}. Similar
results were obtained using Method 3.
\begin{table}
\begin{center}
\begin{tabular}{crrc}
\hline
Peak & $X$         & $Y$              & Local $\ln\mathcal{Z}$ \\
\hline
1 & $-0.400 \pm 0.002$ & $-0.400 \pm 0.002$ & $-9.544 \pm 0.162$ \\
2 & $-0.350 \pm 0.002$ & $ 0.200 \pm 0.002$ & $-8.524 \pm 0.161$ \\
3 & $-0.209 \pm 0.052$ & $ 0.154 \pm 0.041$ & $-6.597 \pm 0.137$ \\
4 & $ 0.100 \pm 0.004$ & $-0.150 \pm 0.004$ & $-7.645 \pm 0.141$ \\
5 & $ 0.449 \pm 0.011$ & $ 0.100 \pm 0.011$ & $-5.689 \pm 0.117$ \\
\hline
\end{tabular}
\caption{The inferred mean values of $X$ and $Y$ and the local evidence
for each Gaussian peak in Toy model 1a using Method 2 (clustered
ellipsoidal sampling).}
\label{tab:Ex1_ces}
\end{center}
\end{table}

In real applications the parameter space is usually of higher
dimension and different modes of the posterior may vary in amplitude
by more than an order of magnitude. To investigate this situation, we
also considered a modified problem in which three 10-dimensional
Gaussians are placed randomly in the unit hypercube $[0,1]$ and have
amplitudes differing by two orders of magnitude. We also make one of
the Gaussians elongated. The analytical local log-evidence values and
those found by applying Method 2 (without sub-clustering) and Method 3
are shown in Table~\ref{tab:Ex1_2}. We used $N=600$ live points with
both of our methods.
\begin{table}
\begin{center}
\begin{tabular}{cccc}
\hline
Toy model 1b & Real Value & Method 2  & Method 3 \\
\hline
$\ln\mathcal{Z}$        &4.66&4.47$\pm$0.20&4.52$\pm$0.20\\
local $\ln\mathcal{Z}_1$&4.61&4.38$\pm$0.20&4.40$\pm$0.21\\
local $\ln\mathcal{Z}_2$&1.78&1.99$\pm$0.21&2.15$\pm$0.21\\
local $\ln\mathcal{Z}_3$&0.00&0.09$\pm$0.20&0.09$\pm$0.20\\
$N_{\rm like}$  &       &130,529&699,778\\
\hline
\end{tabular}
\caption{The true and estimated global log-evidence, local
log-evidence and number of likelihood evaluations required in
analysing Toy model 1b using Method 2 (clustered ellipsoidal sampling)
and Method 3 (Metropolis sampling).}
\label{tab:Ex1_2}
\end{center}
\end{table}

We see that both methods detected all 3 Gaussians and calculated their
evidence values with reasonable accuracy within the estimated
uncertainties. Method 2 required $\sim 4$ times fewer likelihood
calculations than Method 3, since in this problem the ellipsoidal
methods can still achieve very high efficiency (28 per cent), while
the efficiency of the Metropolis method remains constant (5 per cent) as
discussed in Sec.~\ref{metronest}.

\subsection{Toy model 2}
\label{ExII}

We now illustrate the capabilities of our methods in sampling from a
posterior containing multiple modes with pronounced (curving)
degeneracies in high dimensions. Our toy problem is based on that
investigated by \citet{Allanach}, but we extend it to more than two
dimensions.

\begin{figure}
\begin{center}
\includegraphics[width=0.8\columnwidth]{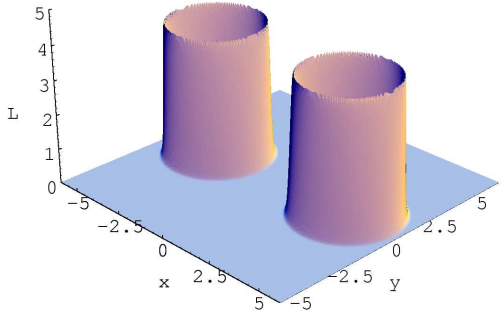}
\caption{Toy model 2: a two-dimensional example of the
likelihood function defined in (\ref{eq:gshellL}) and (\ref{eq:gshell}).}
\label{fig:Ex3_1}
\end{center}
\end{figure}

The likelihood function is defined as,
\begin{equation}
L(\btheta)={\rm circ}(\btheta;\bmath{c}_1,r_1,w_1)
+{\rm circ}(\btheta;\bmath{c}_2,r_2,w_2),
\label{eq:gshellL}
\end{equation}
where
\begin{equation}
{\rm circ}(\btheta;\bmath{c},r,w)=\frac{1}{\sqrt{2\pi
w^2}}\exp\left[-\frac{(\left|\btheta-\bmath{c}\right|-r)^2}{2w^2}\right].
\label{eq:gshell}
\end{equation}
In 2-dimensions, this toy distribution represents two well separated
rings, centred on the points $\bmath{c}_1$ and $\bmath{c}_2$
respectively, each of radius $r$ and with a Gaussian radial profile of
width $w$ (see Fig.~\ref{fig:Ex3_1}).
With a sufficiently small $w$ value, this distribution is
representative of the likelihood functions one might encounter in
analysing forthcoming particle physics experiments in the context of
beyond-the-Standard-Model paradigms; in such models the bulk of the
probability lies within thin sheets or hypersurfaces through the full
parameter space.

We investigate the above distribution up to a $100$-dimensional
parameter space $\btheta$. In all cases, the centers of the two rings
are separated by $7$ units in the parameter space, and we take
$w_1=w_2=0.1$ and $r_1=r_2=2$. We make $r_1$ and $r_2$ equal, since in
higher dimensions any slight difference between these two would result
in a vast difference between the volumes occupied by the rings and
consequently the ring with the smaller $r$ value would occupy
vanishingly small probability volume making its detection almost
impossible. It should also be noted that setting $w=0.1$ means the
rings have an extremely narrow Gaussian profile and hence they
represent an `optimally difficult' problem for our ellipsoidal nested
sampling algorithms, even with sub-clustering, since many tiny
ellipsoids are required to obtain a sufficiently accurate
representation of the iso-likelihood surfaces. For the two-dimensional
case, with the parameters described above, the likelihood function is
that shown in Fig.~\ref{fig:Ex3_1}.

Sampling from such a highly non-Gaussian and curved distribution can
be very difficult and inefficient, especially in higher dimensions. In
such problems a re-parameterization is usually performed to transform
the distribution into one that is geometrically simpler (see
e.g. \citet{Dunkley} and \citet{Verde}), but such approaches are
generally only feasible in low-dimensional problems. In general, in
$D$ dimensions, the transformations usually employed introduce $D-1$
additional curvature parameters and hence become rather inconvenient.
Here, we choose not to attempt a re-parameterization, but instead
sample directly from the distribution.

\begin{figure}
\begin{center}
\includegraphics[width=0.9\columnwidth]{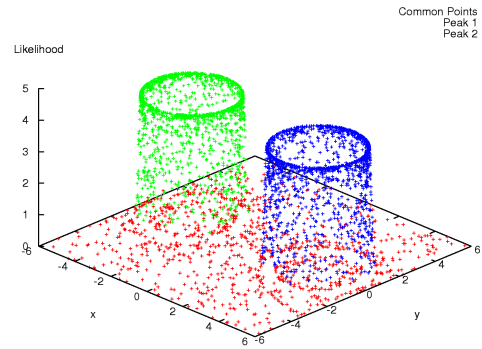}
\caption{Toy Model 2: a two-dimensional posterior consisting of two
rings with narrow Gaussian profiles as defined in
equation~\ref{eq:gshell}. The dots denote the set of live points at
each successive likelihood level in the nested sampling algorithm
using Method 2 (with sub-clustering).}
\label{fig:ex2_2b}
\end{center}
\end{figure}

\begin{table*}
\begin{center}
\begin{tabular}{rrrrrrrrr}
\hline
&\multicolumn{2}{|c|}{Analytical}&\multicolumn{3}{|c|}{Method 2 (with
sub-clustering)}&\multicolumn{3}{|c|}{Method 3}\\
$D$&$\ln\mathcal{Z}$&local $\ln\mathcal{Z}$&$\ln\mathcal{Z}$&local
$\ln\mathcal{Z}_1$&local $\ln\mathcal{Z}_2$&$\ln\mathcal{Z}$&local $\ln\mathcal{Z}_1$&local $\ln\mathcal{Z}_2$\\
\hline
$2$&$-1.75$&$-2.44$&$-1.71\pm0.08$&$-2.41\pm0.09$&$-2.40\pm0.09$&$-1.63\pm0.08$&$-2.35\pm0.09$&$-2.31\pm0.09$\\
$5$&$-5.67$&$-6.36$&$-5.78\pm0.13$&$-6.49\pm0.14$&$-6.46\pm0.14$&$-5.69\pm0.13$&$-6.35\pm0.13$&$-6.41\pm0.14$\\
$10$&$-14.59$&$-15.28$&$-14.50\pm0.20$&$-15.26\pm0.20$&$-15.13\pm0.20$&$-14.31\pm0.19$&$-15.01\pm0.20$&$14.96\pm0.20$\\
$20$&$-36.09$&$-36.78$&$-35.57\pm0.30$&$-36.23\pm0.30$&$-36.20\pm0.30$&$-36.22\pm0.30$&$-36.77\pm0.31$&$-37.09\pm0.31$\\
$30$&$-60.13$&$-60.82$&$$&$$&$$&$-60.49\pm0.39$&$-61.69\pm0.39$&$-60.85\pm0.39$\\
$50$&$-112.42$&$-113.11$&$$&$$&$$&$-112.27\pm0.53$&$-112.61\pm0.53$&$-113.53\pm0.53$\\
$70$&$-168.16$&$-168.86$&$$&$$&$$&$-167.71\pm0.64$&$-167.98\pm0.64$&$-169.32\pm0.65$\\
$100$&$-255.62$&$-256.32$&$$&$$&$$&$-253.72\pm0.78$&$-254.16\pm0.78$&$-254.77\pm0.78$\\
\hline
\end{tabular}
\caption{The true and estimated global and local $\log\mathcal{Z}$ for
toy model 3, as a function of the dimensions $D$ of the parameter
space, using Method 2 (with sub-clustering) and Method 3.}
\label{tab:ex2.1}
\end{center}
\end{table*}

Applying the ellipsoidal nested sampling approaches (methods 1 and 2)
to this problem without using the sub-clustering modification would
result in highly inefficient sampling as the enclosing ellipsoid would
represent an extremely poor approximation to the ring. Thus, for this
problem, we use Method 2 with sub-clustering and Method 3 (Metropolis
nested sampling).  We use $400$ live points in both algorithms. The
sampling statistics are listed in Table~\ref{tab:ex2.1} and
Table~\ref{tab:ex2.2} respectively.  The 2-dimensional sampling
results using Method 2 (with sub-clustering) are also illustrated in
Fig.~\ref{fig:ex2_2b}, in which the set of live points at each
successive likelihood level is plotted; similar results are obtained
using Method 3.

We see that both methods produce reliable estimates of the global and
local evidences as the dimension $D$ of the parameter space increases.
As seen in Table~\ref{tab:ex2.2}, however, the efficieny of Method 2,
even with sub-clustering, drops significantly with increasing
dimensionality. As a result, we do not explore the problem with method
2 for dimensions greater than $D=20$. This drop in efficiency is
caused by (a) in higher dimensions even a small region of an ellipsoid
that lies outside the true iso-likelihood contour occupies a large
volume and hence results in a drop in sampling efficiency; and (b) in
$D$ dimensions, the minimum number of points in an ellipsoid can be
$(D+1)$, as discussed in Sec.~\ref{method:degeneracies}, and
consequently with a given number of live points, the number of
sub-clusters decreases with increasing dimensionality, resulting in a
poor approximation to the highly curved iso-likelihood contour.
Nonetheless, Method 3 is capable of obtaining evidence estimates with
reasonable efficiency up to $D=100$, and should continue to operate
effectively at even higher dimensionality.

\begin{table}
\begin{center}
\begin{tabular}{rrrrr}
\hline
&\multicolumn{2}{|c|}{Method 2 (with sub-clustering)}&\multicolumn{2}{|c|}{Method 3}\\
$D$ & $N_{\rm like}$ & Efficiency & $N_{\rm like}$ & Efficiency\\
\hline
$2$&$27,658$&$15.98\%$&$76,993$&$6.07\%$\\
$5$&$69,094$&$9.57\%$&$106,015$&$6.17\%$\\
$10$&$579,208$&$1.82\%$&$178,882$&$5.75\%$\\
$20$&$43,093,230$&$0.05\%$&$391,113$&$5.31\%$\\
$30$&$$&$$&$572,542$&$5.13\%$\\
$50$&$$&$$&$1,141,891$&$4.95\%$\\
$70$&$$&$$&$1,763,253$&$4.63\%$\\
$100$&$$&$$&$3,007,889$&$4.45\%$\\
\hline
\end{tabular}
\caption{The number of likelihood evaluations and sampling efficiency
 for Method 2 (with sub-clustering) and Method 3 when applied to toy
 model 3, as a function of the dimension $D$ of the parameter space}
\label{tab:ex2.2}
\end{center}
\end{table}

\section{Bayesian object detection}
\label{implementation:ExampleIII}
We now consider how our multimodal nested sampling approaches may be
used to address the difficult problem of detecting and characterizing
discrete objects hidden in some background noise.  A Bayesian approach
to this problem in an astrophysical context was first presented by
Hobson \& McLachlan (2003; hereinafter HM03), and our general
framework follows this closely. For brevity, we will consider our data
vector $\boldsymbol{D}$ to denote the pixel values in a single image in
which we wish to search for discrete objects, although
$\boldsymbol{D}$ could equally well represent the Fourier coefficients
of the image, or coefficients in some other basis.

\subsection{Discrete objects in background}
\label{Ex3:discrete_obj}

Let us suppose we are interested in detecting and characterising some
set of (two-dimensional) discrete objects, each of which is described
by a template $\tau(\bmath{x};\bmath{a})$, which is parametrised in
terms of a set of parameters $\bmath{a}$ that might typically denote
(collectively) the position $(X,Y)$ of the object, its amplitude $A$
and some measure $R$ of its spatial extent. In particular, in this
example we will assume circularly-symmetric Gaussian-shaped objects
defined by
\begin{equation}
\tau(\bmath{x};\bmath{a})=A\exp
\left[-\frac{(x-X)^2+(y-Y)^2}{2R^2}\right],
\label{objdef}
\end{equation}
so that $\bmath{a} = \{X,Y,A,R\}$. If $N_{\rm obj}$ such
objects are present and the contribution of each object to the data 
is additive, we may write
\begin{equation}
\bmath{D} = \bmath{n}+ \sum_{k=1}^{N_{\rm obj}} \bmath{s}(\bmath{a}_k),
\end{equation}
where $\bmath{s}(\bmath{a}_k)$ denotes the contribution to the data
from the $k$th discrete object and $\bmath{n}$ denotes the generalised
`noise' contribution to the data from other `background' emission and
instrumental noise. Clearly, we wish to use the data $\bmath{D}$ to
place constraints on the values of the unknown parameters $N_{\rm
obj}$ and $\bmath{a}_k$ $(k=1,\ldots,N_{\rm obj})$.

\subsection{Simulated data}
\label{Ex3}

\begin{figure*}
\centerline{
\includegraphics[width=0.6\textwidth]{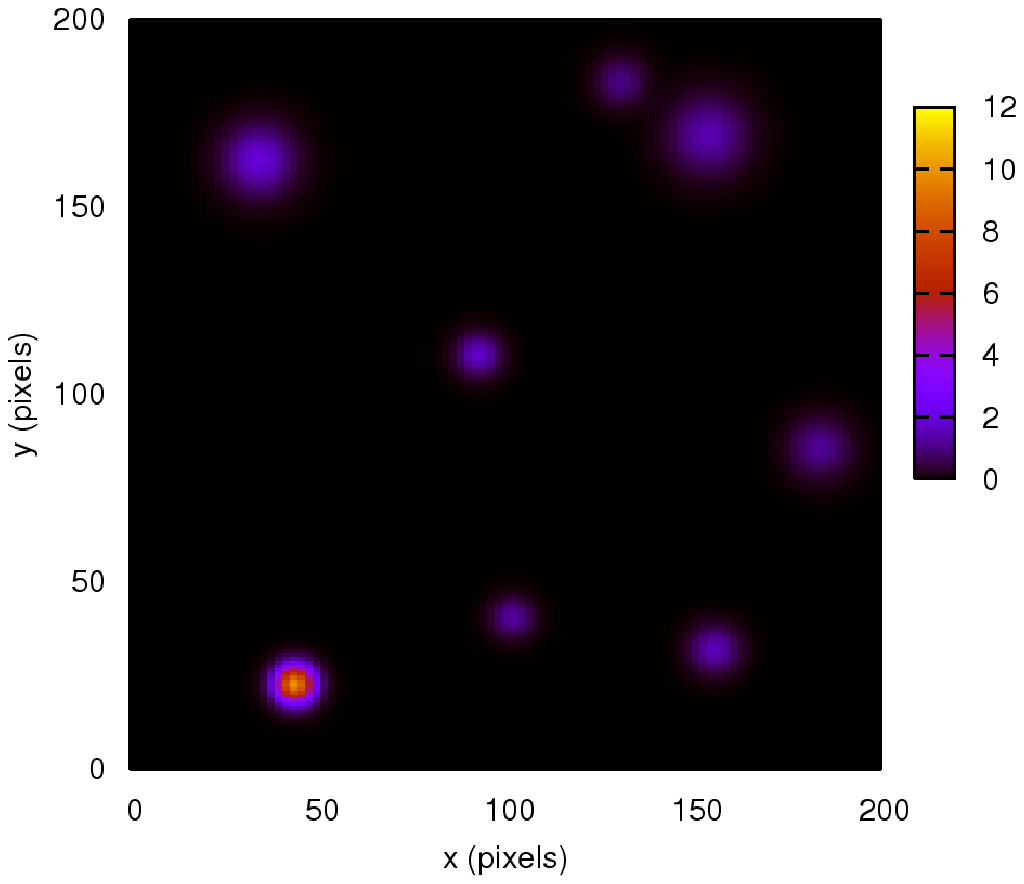}
\includegraphics[width=0.6\textwidth]{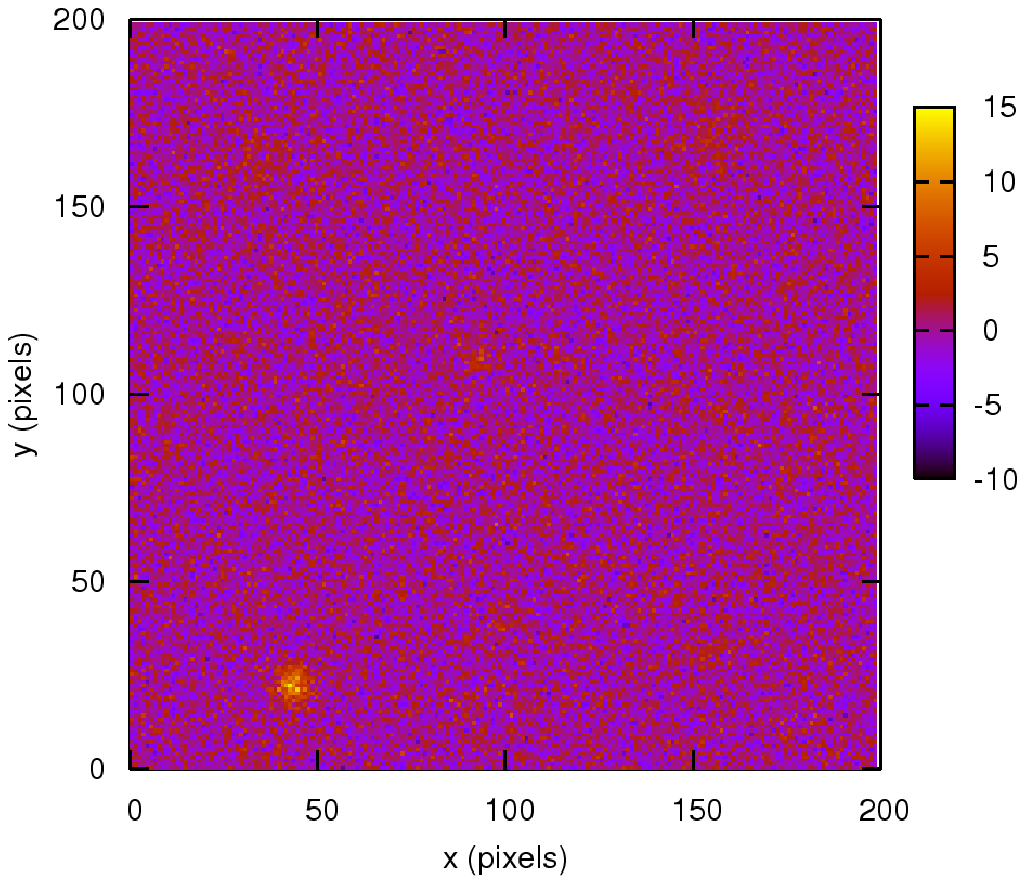}
      }
\caption{The toy model discussed in Sec.~\ref{Ex3}. The
$200\times 200$ pixel test image (left panel) contains 8 Gaussian 
objects of varying widths and amplitudes; the parameters
$X_{k}$, $Y_{k}$, $A_{k}$ and $R_{k}$ for each object are listed in
Table~\ref{tab:Ex3.1}. The right panel shows the corresponding data map
with independent Gaussian noise added with an rms of 2 units.} \label{f71}
\end{figure*}

Our underlying model and simulated data are shown in Fig.~\ref{f71},
and are similar to the example considered by HM03.  The left panel
shows the $200\times 200$ pixel test image, which contains 8 Gaussian
objects described by eq.~(\ref{objdef}) with the parameters $X_{k}$,
$Y_{k}$, $A_{k}$ and $R_{k}$ $(k=1,...,8)$ listed in
Table~\ref{tab:Ex3.1}. The $X$ and $Y$ coordinates are drawn
independently from the uniform distribution $U(0,200)$.  Similarly,
the amplitude $A$ and size $R$ of each object are drawn independently
from the uniform distributions $U(1,2)$ and $U(3,7)$ respectively. We multiply the
amplitude of the first object by $10$ to to see how senstive our nested sampling methods 
are to this order of magnitude difference in amplitudes. The
simulated data map is created by adding independent Gaussian pixel
noise with an rms of 2 units. This corresponds to a signal-to-noise
ratio 0.5-1 as compared to the peak amplitude of each object (ignoring the first object). 
It can be seen from the figure that with this level of noise, apart from the first object,
only a few objects are (barely) visible with the naked eye and there are certain
areas where the noise conspires to give the impression of an object
where none is present. This toy problem thus presents a considerable
challenge for any object detection algorithm.

\begin{table}
\begin{center}
\begin{tabular}{crrrr}
\hline
Object & $X$  & $Y$ & $A$ & $R$ \\
\hline
1 &  43.71 &  22.91 & 10.54 & 3.34\\
2 & 101.62 &  40.60 &  1.37 & 3.40\\
3 &  92.63 & 110.56 &  1.81 & 3.66\\
4 & 183.60 &  85.90 &  1.23 & 5.06\\
5 &  34.12 & 162.54 &  1.95 & 6.02\\
6 & 153.87 & 169.18 &  1.06 & 6.61\\
7 & 155.54 &  32.14 &  1.46 & 4.05\\
8 & 130.56 & 183.48 &  1.63 & 4.11\\
\hline
\end{tabular}
\caption{The parameters $X_{k}$, $Y_{k}$, $A_{k}$ and $R_{k}$
$(k=1,...,8)$ defining the Gaussian shaped objects in Fig.~\ref{f71}.}
\label{tab:Ex3.1}
\end{center}
\end{table}

\subsection{Defining the posterior distribution}

As discussed in HM03, in analysing the above simulated data map the
Bayesian purist would attempt to infer simultaneously the full set of
parameters $\mathbf{\Theta} \equiv (N_{\rm obj},
\bmath{a}_1,\bmath{a}_2,\ldots, \bmath{a}_{N_{\rm obj}})$. The crucial
complication inherent to this approach is that the length of the
parameter vector $\mathbf{\Theta}$ is variable, since it depends on
the unknown value $N_{\rm obj}$. Thus any sampling based approach must
be able to move between spaces of different dimensionality, and such
techniques are investigated in HM03.

An alternative approach, also discussed by HM03, is simply to set
$N_{\rm obj}=1$. In other words, the model for the data consists of
just a single object and so the full parameter space under
consideration is $\bmath{a}=\{X,Y,A,R\}$, which is fixed and
only 4-dimensional.  Although fixing $N_{\rm obj}=1$, it is
important to understand that this does {\em not} restrict us to
detecting just a single object in the data map.  Indeed, by modelling
the data in this way, we would expect the posterior distribution to
possess numerous local maxima in the 4-dimensional parameter space,
each corresponding to the location in this space of one of the
objects present in the image. HM03 show this vastly simplified
approach is indeed reliable when the objects of interest are spatially
well-separated, and so for illustration we adopt this method here.

In this case, if the background `noise' $\bmath{n}$ is a statistically
homogeneous Gaussian random field with covariance matrix $\mathbf{N} = 
\langle \bmath{n}\bmath{n}^{\rm t} \rangle$, then the likelihood
function takes the form
\begin{equation}
L(\bmath{a})=\frac{\exp\left\{
-\frac{1}{2}\left[\bmath{D}-\bmath{s}(\bmath{a})\right]^{\rm t}
\mathbf{N}^{-1}\left[\bmath{D}-\bmath{s}(\bmath{a})\right]\right\} 
}{\left(2\pi\right)^{N_{\rm pix}/2}\left|\mathbf{N}\right|^{1/2}}.
\label{eq:22}
\end{equation}
In our simple problem the background is just independent pixel noise, so
$\mathbf{N} = \sigma^2 \mathbf{I}$, where $\sigma$ is the noise rms.
The prior on the parameters is assumed to be separable, so that
\begin{equation}
\pi(\bmath{a})=\pi(X)\pi(Y)\pi(A)\pi(R).
\end{equation}
The priors on $X$ and $Y$ are taken to be the uniform distribution
$U(0,200)$, whereas the priors on $A$ and $R$ are taken as the uniform
distributions $U(1,12.5)$ and $U(2,9)$ respectively.

The problem of object identification and characterization then
reduces to sampling from the (unnormalised) posterior to infer
parameter values and calculating the `local' Bayesian evidence for each
detected object to assess the probability that it is indeed real.
In the most straightforward approach, the two competing models 
between which we must select are $H_0 = $ `the detected object is
fake $(A=0)$' and $H_1 = $ `the detected object is real $(A > 0)$'. One
could, of course, consider alternative definitions of these
hypotheses, such as setting $H_0$: $A \le A_{\rm lim}$ and $H_1$: $A
> A_{\rm lim}$, where $A_{\rm lim}$ is some (non-zero) cut-off
value below which one is not interested in the identified object.
 
\subsection{Results}
\label{Ex3_res}

\begin{figure}
 \begin{center}
    	\subfigure[]{
          \includegraphics[width=0.82\columnwidth]{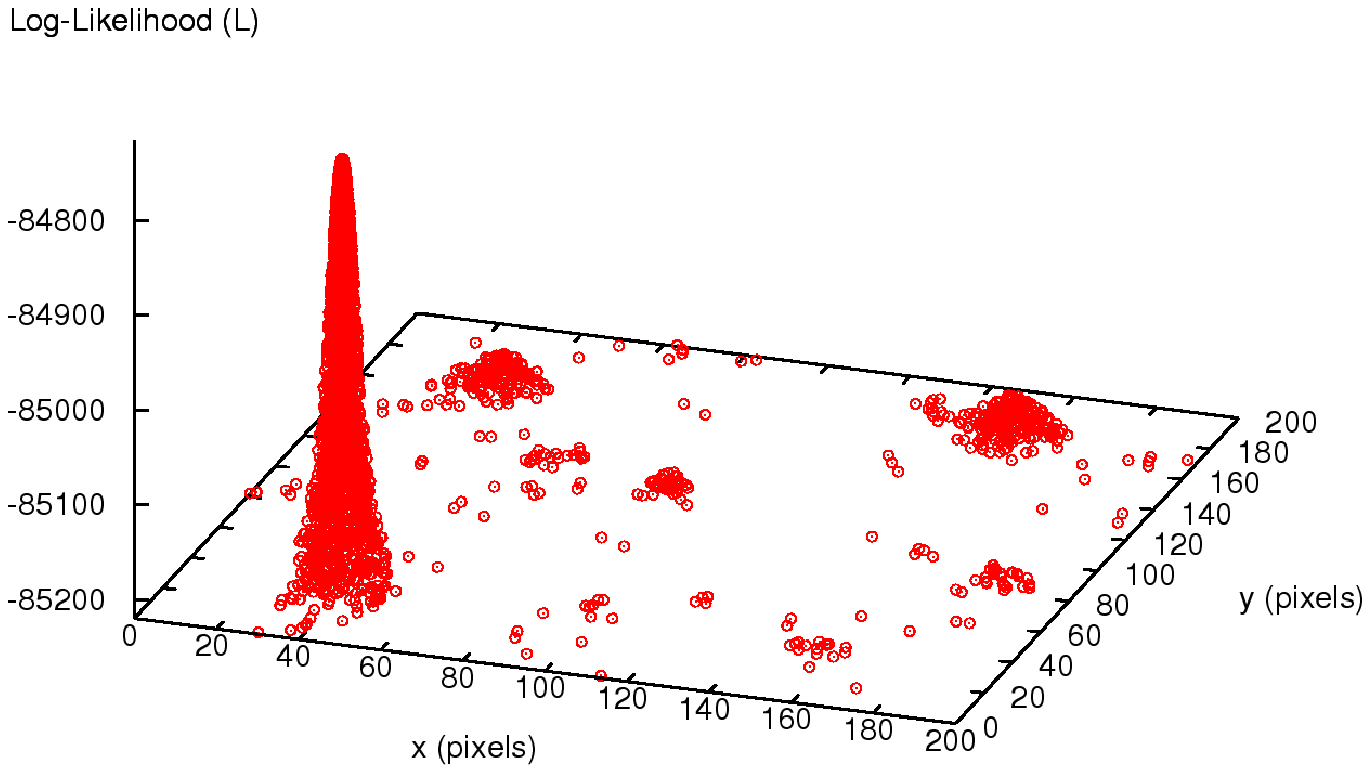}}
	\subfigure[]{
          \includegraphics[width=0.82\columnwidth]{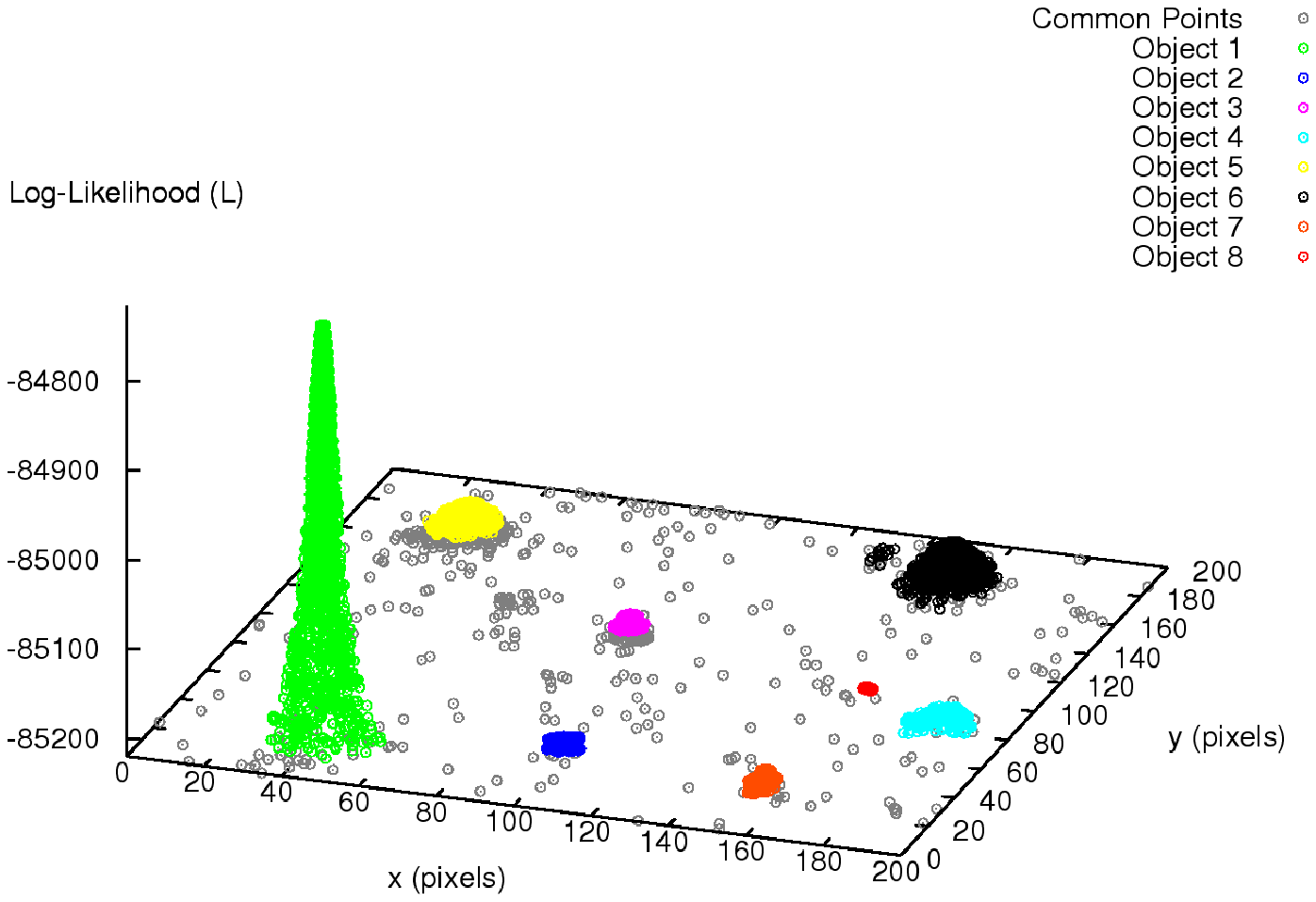}}
      \subfigure[]{
          \includegraphics[width=0.82\columnwidth]{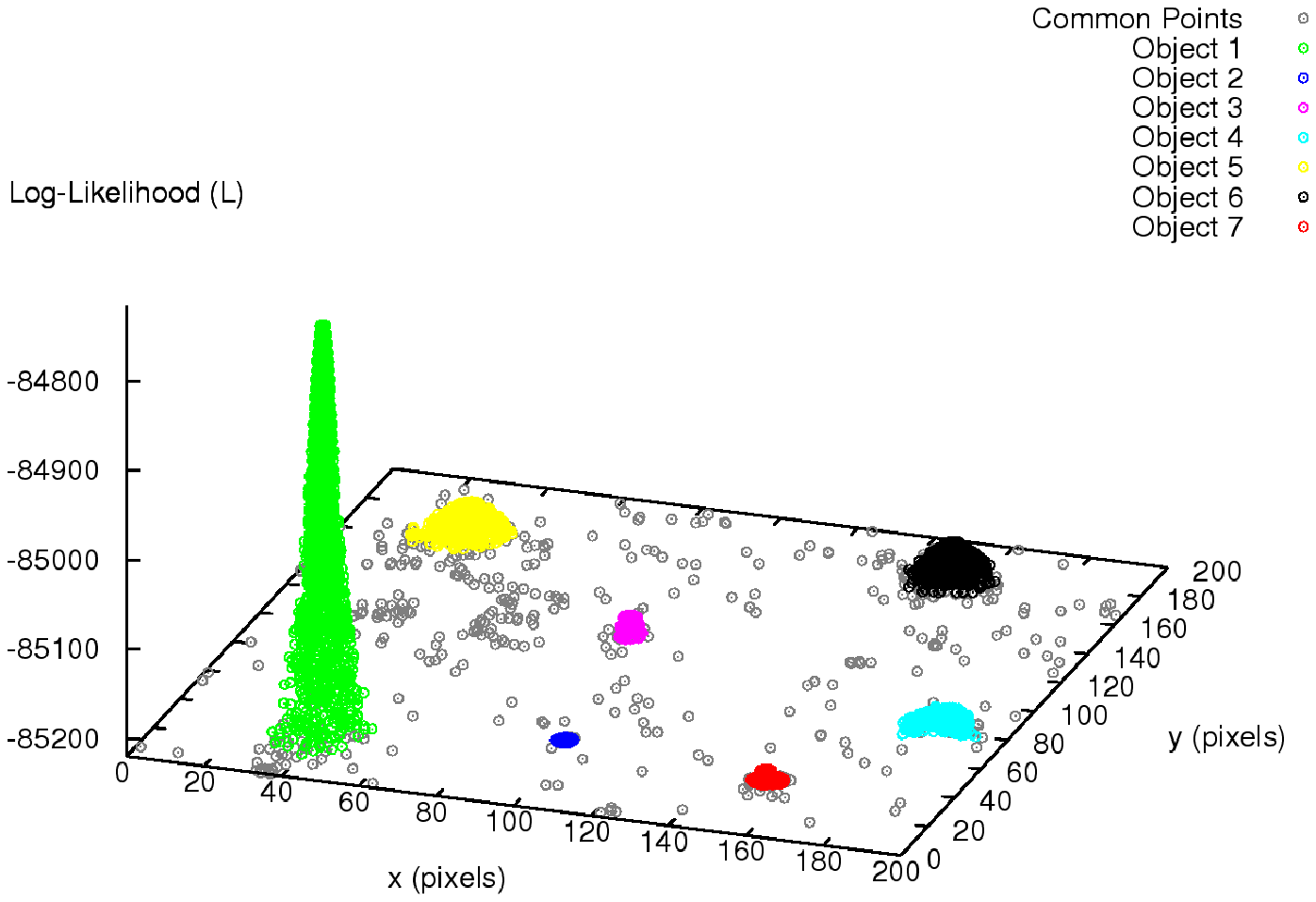}}
\caption{The set of live points, projected into the $(X,Y)$-subspace,
at each successive likelihood level in the nested sampling in the
analysis of the data map in Fig.~\ref{f71}(right panel) using: (a)
Method 1 (no sub-clustering); (b) Method 2 (no sub-clustering); and
(c) Method 3. In (b) and (c) the different colours denote points
assigned to isolated clusters as the algorithm progresses.}
\label{fig:Ex3_res}
\end{center}
\end{figure}

Since Bayesian object detection is of such interest, we analyse this
problem using methods 1, 2 and 3. For methods 1 and 2, do not use
sub-clustering, since the posterior peaks are not expected to exhibit
pronounced (curving) degeneracies. We use 400 live points with method
1 and 300 with methods 2 and 3. In methods 1 and 2, the initial
enlargement factor was set to $f_0=0.3$.

In Fig.~\ref{fig:Ex3_res} we plot the live points, projected into the
$(X,Y)$-subspace, at each successive likelihood level in the nested
sampling algorithm (above an arbitrary base level) for each
method. For the methods 2 and 3 results, plotted in panel (b) and (c)
respectively, the different colours denote points assigned to isolated
clusters as the algorithm progresses; we note that the base likelihood
level used in the figure was chosen to lie slightly below that at
which the individual clusters of points separate out. We see from the
figure, that all three approaches have successfully sampled from this
highly multimodal posterior distribution.  As discussed in HM03, this
represents a very difficult problem for traditional MCMC methods, and
illustrates the clear advantages of our methods. In detail, the figure
shows that samples are concentrated in 8 main areas. Comparison with
Fig.~\ref{f71} shows that 7 of these regions do indeed correspond to
the locations of the real objects (one being a combination of two real
objects), whereas the remaining cluster corresponds to a
`conspiration' of the background noise field. The CPU time required
for Method 1 was only $\sim 5$ minutes on a single Itanium 2
(Madison) processor of the {\sc Cosmos} supercomputer; each processor
has a clock speed of 1.3~GHz, a 3Mb L3 cache and a peak performance of
5.2 Gflops.

\begin{table}
\begin{center}
\begin{tabular}{cccc}
\hline
                  & Method 1           & Method 2   & Method 3   \\
                  & (no sub-clustering)& (no sub-clustering) &            \\
\hline
$\ln\mathcal{Z}$ & $-84765.63$        & $-84765.41$   & $-84765.45$  \\
Error             &  0.20              &   0.24     &  0.24      \\  
$N_{\rm like}$    &  55,521            & 74,668     & 478,557    \\
\hline
\end{tabular}
\caption{Summary of the global evidence estimates for toy model 3 and
the number of likelihood evaluations required using different sampling
methods. The `null' log-evidence for the model in which no object is
present is $-85219.44$.}
\label{tab:Ex3_res}
\end{center}
\end{table}

\begin{table*}
\begin{center}
\begin{tabular}{ccrrrr}
\hline
Cluster & local $\ln\mathcal{Z}$ & $X$  & $Y$ & $A$ & $R$ \\
\hline
1 & $-84765.41 \pm 0.24$ &  43.82 $\pm$ 0.05 &  23.17 $\pm$ 0.05 & 10.33 $\pm$ 0.15 & 3.36 $\pm$ 0.03\\
2 & $-85219.61 \pm 0.19$ & 100.10 $\pm$ 0.26 &  40.55 $\pm$ 0.32 &  1.93 $\pm$ 0.16 & 2.88 $\pm$ 0.15\\
3 & $-85201.61 \pm 0.21$ &  92.82 $\pm$ 0.14 & 110.17 $\pm$ 0.16 &  3.77 $\pm$ 0.26 & 2.42 $\pm$ 0.13\\
4 & $-85220.34 \pm 0.19$ & 182.33 $\pm$ 0.48 &  85.85 $\pm$ 0.43 &  1.11 $\pm$ 0.07 & 4.85 $\pm$ 0.30\\
5 & $-85194.16 \pm 0.19$ &  33.96 $\pm$ 0.36 & 161.50 $\pm$ 0.35 &  1.56 $\pm$ 0.09 & 6.28 $\pm$ 0.29\\
6 & $-85185.91 \pm 0.19$ & 155.21 $\pm$ 0.31 & 169.76 $\pm$ 0.32 &  1.69 $\pm$ 0.09 & 6.48 $\pm$ 0.24\\
7 & $-85216.31 \pm 0.19$ & 154.87 $\pm$ 0.32 &  31.59 $\pm$ 0.22 &  1.98 $\pm$ 0.17 & 3.16 $\pm$ 0.20\\
8 & $-85223.57 \pm 0.21$ & 158.12 $\pm$ 0.17 &  96.17 $\pm$ 0.19 &  2.02 $\pm$ 0.10 & 2.15 $\pm$ 0.09\\
\hline
\end{tabular}
\caption{The mean and standard deviation of the evidence and inferred object parameters 
$X_{k}$, $Y_{k}$, $A_{k}$ and $R_{k}$ for toy model 4 using Method 2.}
\label{tab:Ex3.2}
\end{center}
\end{table*}

The global evidence results are summarised in Table~\ref{tab:Ex3_res}.
We see that all three approaches yield consistent values within the
estimated uncertainties, which is very encouraging given their
considerable algorithmic differences. 

We note, in particular, that Method 3 required more than 6 times the
number of likelihood evaluations as compared to the ellipsoidal
methods. This is to be expected given the non-degenerate shape of the
posterior modes and the low-dimensionality of this problem. The global
evidence value of $ \sim -84765$ may be interpreted as corresponding
to the model $H_1 = $ `there is a real object somewhere in the
image'. Comparing this with the `null' evidence value $\sim -85219$
for $H_0 = $ `there is no real object in the image', we see that $H_1$
is strongly favoured, with a log-evidence difference of $\Delta
\ln\mathcal{Z} \sim 454$.

In object detection, however, one is more interested in whether or not
to believe the individual objects identified. As discussed in
Sections~\ref{method:clustered} and \ref{method:levidence}, using
Method 2 and Method 3, samples belonging to each
identified mode can be separated and local evidences and posterior
inferences calculated. In Table~\ref{tab:Ex3.2}, for each separate
cluster of points, we list the mean and standard error of the inferred
object parameters and the local $\log$-evidence obtained using method
2; similar results are obtained from Method 3.  Considering first the
local evidences and comparing them with the `null' evidence of
$-85219.44$, we see that all the identified clusters should be
considered as real detections, except for cluster 8. Comparing the
derived object parameters with the inputs listed in
Table~\ref{tab:Ex3.1}, we see that this conclusion is indeed correct.
Moreover, for the 7 remaining clusters, we see that the derived
parameter values for each object are consisent with the true values.

It is worth noting, however, that cluster 6 does in fact correspond to
the real objects 6 and 8, as listed in Table~\ref{tab:Ex3.1}.  This
occurs because object 8 lies very close to object 6, but has a much
lower amplitude. Although one can see a separate peak in the posterior
at the location of object 8 in Fig.~\ref{fig:Ex3_res}(c) (indeed this
is visible in all three panels), Method 2 was not able to identify a
separate, isolated cluster for this object.  Thus, one drawback of
clustered ellipsoidal sampling method is that it may not identify all
objects in a set lying very close together and with very different
amplitudes. This problem can be overcome by increasing the number of
objects assumed in the model from $N_{\rm obj}=1$ to some appropriate
larger value, but we shall not explore this further here. It should be
noted however, that failure to separate out every real object
has no impact on the accuracy of the estimated {\em global} evidence,
since the algorithm still samples from a region that includes all the
objects.

\section{Discussion and conclusions}

In this paper, we have presented various methods that allow the
application of the nested sampling algorithm \citep{Skilling} to
general distributions, particular those with multiple modes and/or
pronounced (curving) degeneracies. As a result, we have produced a
general Monte Carlo technique capable of calculating Bayesian evidence
values and producing posterior inferences in an efficient and robust
manner. As such, our methods provide a viable alternative to MCMC
techniques for performing Bayesian analyses of astronomical data sets.
Moreover, in the analysis of a set of toy problems, we demonstrate
that our methods are capable of sampling effectively from posterior
distributions that have traditionally caused problems for MCMC
approaches. Of particular interest is the excellent performance of our
methods in Bayesian object detection and validation, but
our approaches should provide advantages in all areas of Bayesian
astronomical data analysis.

A critical analysis of Bayesian methods and MCMC sampling has recently
been presented by \citet{Bryan}, who advocate a frequentist
approach to cosmological parameter estimation from the CMB power
spectrum.  While we refute wholeheartedly their criticisms of Bayesian
methods {\em per se}, we do have sympathy with their assessment of
MCMC methods as a poor means of performing a Bayesian inference. In
particular, \citet{Bryan} note that for MCMC sampling methods ``if a
posterior is comprised by two narrow, spatially separated Gaussians,
then the probability of transition from one Gaussian to the other will
be vanishingly small. Thus, after the chain has rattled around in one
of the peaks for a while, it will appear that the chain has converged;
however, after some finite amount of time, the chain will suddenly
jump to the other peak, revealing that the initial indications of
convergence were incorrect.'' They also go on to point out that MCMC
methods often require considerable tuning of the proposal distribution
to sample efficiently, and that by their very nature MCMC samples are
concentrated at the peak(s) of the posterior distribution often
leading to underestimation of confidence intervals when time allows
only relatively few samples to be taken. We believe our multimodal
nested sampling algorithms address all these criticisms. Perhaps of
most relevance is the claim by \citet{Bryan} that their analysis of the
1-year WMAP \citep{Bennett} identifies two distinct regions of high
posterior probability in the cosmological parameter space.  Such
multimodality suggests that our methods will be extremely useful in
analysing WMAP data and we will investigate this in a forthcoming publication.

The progress of our multimodel nested sampling algorithms based on ellipsoidal sampling
(methods 1 and 2) is controlled by 
three main parameters: (i) the number of live points $N$; (ii) the initial 
enlargement factor $f_0$; and (iii) the rate $\alpha$ at which the 
enlargement factor decreases with decreasing prior volume. The approach based on
Metropolis nested sampling (Method 3) depends only on $N$. These values 
can be chosen quite easily as outlined below and the performance of the algorithm   
is relatively insensitive to them. First, $N$ should be large enough that, 
in the initial sampling from the full prior space, there is a high 
probability that at least one point lies in the `basin of attraction' of 
each mode of the posterior. In later iterations, live points will then 
tend to populate these modes. Thus, as a rule of thumb, one should take 
$N \ga V_\pi/V_{\rm mode}$, where $V_{\rm mode}$ is (an estimate of) the 
volume of the posterior mode containing the smallest probability volume 
(of interest) and $V_\pi$ is the volume of the full prior space. It should 
be remembered, of course, that $N$ must always exceed the dimensionality 
$D$ of the parameter space. Second, $f_0$ should usually be set in the 
range 0--0.5. At the initial stages, a large value of $f_0$ is required to take
into account the error in approximating a large prior volume with ellipsoids
contructed from limited number of live points. Typically, a value of 
$f_0 \sim 0.3$ should suffice for $N \sim 300$.
The dynamic enlargement factor $f_{i,k}$ gradually goes down with decreasing
prior volume and consequently, increasing the sampling efficiency as discussed in
\ref{improvements:def}. Third, $\alpha$ should be set in the 
range 0--1, but typically a value of $\alpha \sim 0.2$ is appropriate for 
most problems. The algorithm also depends on a few additional parameters, 
such as the number of previous iterations to consider when matching clusters 
in Method 2 (see Section~\ref{method:clustered}), and the number of points 
shared between sub-clusters when sampling from degeneracies (see 
Section~\ref{method:degeneracies}), but there is generally no need to change 
them from their default values.

Looking forward to the further development of our approach, we note
that the new methods presented in this paper operate by providing an
efficient means for performing the key step at each iteration of a
nested sampling process, namely drawing a point from the prior within
the hard constraint that its likelihood is greater than that of the
previous discarded point. In particular, we build on the ellipsoidal
sampling approaches previously suggested by \citet{Mukherjee} and
\citet{Shaw}.  One might, however, consider replacing each hard-edged
ellipsoidal bound by some softer-edged smooth probability
distribution. Such an approach would remove the potential (but
extemely unlikely) problem that some part of the true iso-likelihood
contour may lie outside the union of the ellipsoidal bounds, but it
does bring additional complications. In particular, we explored the
use of multivariate Gaussian distributions defined by the covariance
matrix of the relevant live points, but found that the large tails of
such distributions considerably reduced the sampling efficiency in
higher-dimensional problems. The investigation of alternative
distributions with heavier tails is ongoing. Another difficulty in
using soft-edged distributions is that the method for sampling
consistent from overlapping regions becomes considerably more
complicated, and this too is currently under investigation.

We intend to apply our new multimodal nested sampling methods to a range
of astrophysical data analysis problems in a number of forthcoming papers.
Once we are satisfied that the code performs as anticipated in these test
cases, we plan to make a Fortran library containing our routines
publically available. Anyone wishing to use our code prior to the public
release should contact the authors.

\section*{Acknowledgements}

This work was carried out largely on the {\sc Cosmos} UK National
Cosmology Supercomputer at DAMTP, Cambridge and we thank Stuart Rankin
and Victor Treviso for their computational assistance. We also thank
Keith Grainge, David MacKay, John Skilling, Michael Bridges, Richard Shaw and Andrew
Liddle for extremely helpful discussions. FF is supported by
fellowships from the Cambridge Commonwealth Trust and the Pakistan
Higher Education Commission.

\end{document}